\documentclass[preprint]{aastex62}
\usepackage{amsmath}
\usepackage{mathtools}
\usepackage{xfrac}
\usepackage{color}
\usepackage[normalem]{ulem}

\newcommand{\dd}{{\mathrm d}}
\newcommand{\rmsub}[1]{_\mathrm{#1}}
\newcommand{\molH}{\ensuremath{\mathrm{H_2}}}
\newcommand{\HI}{H\,\textsc{i}}

\graphicspath{{./}{figures/}}

\shorttitle{DIB KT and the Milky Way's spiral structure}
\shortauthors{Tchernyshyov, Peek, Zasowski}

\begin{document}

\title{Kinetic Tomography. II. A second method for mapping the velocity field of the Milky Way Interstellar Medium and a comparison with spiral structure models}

\author[0000-0003-0789-9939]{Kirill Tchernyshyov}
\affiliation{Department of Physics and Astronomy, Johns Hopkins University, 3400 N. Charles Street, Baltimore, MD 21218, USA}
\correspondingauthor{Kirill Tchernyshyov}
\email{ktcherny@gmail.com}

\author[0000-0003-4797-7030]{J. E. G. Peek}
\affiliation{Department of Physics and Astronomy, Johns Hopkins University, 3400 N. Charles Street, Baltimore, MD 21218, USA}
\affiliation{Space Telescope Science Institute, 3700 San Martin Drive, Baltimore, MD 21218, USA}

\author{Gail Zasowski}
\affiliation{Space Telescope Science Institute, 3700 San Martin Drive, Baltimore, MD 21218, USA}
\affiliation{Department of Physics and Astronomy, University of Utah, Salt Lake City, UT 84105, USA}

\begin{abstract}
In this work, we derive a spatially resolved map of the line-of-sight velocity of the interstellar medium and use it, along with a second map of line-of-sight velocity from Paper I of this series, to determine the nature of gaseous spiral structure in the Milky Way.
This map is derived from measurements of the 1.527 $\mu$m diffuse interstellar band (DIB) in stellar spectra from the APOGEE survey and covers the nearest 4-5 kpc of the Northern Galactic plane.
We cross-check this new DIB-based line-of-sight velocity map with the map derived in Paper I and find that they agree.
We then compare these maps with line-of-sight velocity maps derived from simulations of quasi-stationary density wave spiral structure and dynamic, or material, spiral structure in a Milky Way-like galaxy.
While none of the maps derived from these simulations is an exact match to the measured velocity field of the Milky Way, the measurements are more consistent with simulations of dynamic spiral structure.
In the dynamic spiral structure simulation that best matches the measurements, the Perseus spiral arm is being disrupted.
\end{abstract}


\section{Introduction} \label{sec:intro}
There are two main models for how a differentially rotating galaxy can have long-lived spiral structure.
The first is the stationary density wave model \citep[SDW;][]{1964ApJ...140..646L,Shu:2016ij}.
In the SDW model, spiral arms are global oscillatory modes of a stellar or gaseous disk.
The group velocity of a wave does not have equal the velocity of the oscillating matter, so the arms can propagate without winding up.
The second model for long-lived spiral structure is known as the dynamic, the transient and recurrent, or the material spiral structure model \citep{1984ApJ...282...61S}.
Here, we will use the term ``dynamic'' to describe these models.
In the dynamic spiral structure model, the pattern is corotating with the matter.
Individual spiral arms form through a process such as swing amplification, wind up, and dissipate over one to a few Galactic rotation periods \citep{1984ApJ...282...61S,2013ApJ...766...34D}.
If the arm formation process is efficient, these dissipating arms are rapidly replaced, meaning that although individual spiral arms are short-lived, spiral structure in general is long-lived.
It is not known whether the type of spiral structure in most spiral galaxies is SDW or dynamic.
Evidence has been found for both models, sometimes in the same galaxies; see \citet{Shu:2016ij} for examples of evidence in support of the SDW model and \citet{2014PASA...31...35D} in support of the dynamic model.
In this paper, we focus on the spiral structure of the Milky Way.

Since the 1950s, the consensus has been that the Milky Way has spiral arms in has spiral arms in gas, star formation, and young stars \citep{Morgan:1952gm,1954BAN....12..117V}
The gaseous spiral arms are detected as contiguous features in $\ell-v$ diagrams and have been seen in \HI{}-traced neutral gas \citep{1954BAN....12..117V} and CO-traced molecular gas \citep{1980ApJ...239L..53C}.
Arms traced by young stars \citep[e.g.][]{Morgan:1952gm,Xu:2018kg} and star formation regions \citep[e.g.,][]{2014ApJ...783..130R} are detected as contiguous overdensities in space.
Emission from the gaseous arms can be detected out to large distances, including the far side of the Galaxy \citep{Dame_2011_outerspiralarm}, but their positions in space can only be inferred using indirect methods such as the kinematic distance method.
The situation is reversed for the star forming arms -- their positions in space are known directly, allowing measurements of arm properties such as pitch angles to be made, but the necessary observations are not available at large distances from the Sun.
The measurable distribution of stars and star formation in $\ell$, $b$, and $d$ and of gas in $\ell$, $b$, and $v_d$ can fit into the context of either model of spiral structure and cannot decisively distinguish between them.
In this work, we investigate what can be determined about the Milky Way's spiral structure from the velocity field of its interstellar medium (ISM).

These two theories of spiral structure make different predictions for large-scale streaming motions, i.e. spatially coherent deviations from simple rotation.
Spiral structure induces, and is produced by, streaming motions.
These streaming motions should be particularly clear in the velocity field of interstellar matter, which is collisional and hence dynamically cold.
In the SDW model, interstellar matter flows \emph{through} a spiral arm or, equivalently, the overdensity of ISM that is an arm moves through the disk of the galaxy \citep{1969ApJ...158..123R}.
An SDW arm is simultaneously accumulating matter from one side and losing it from the other in a flow that spans the entire length of the arm.
In the dynamic model, gas converges on a spiral arm that is growing and is sheared or blown away (e.g., by stellar and supernova feedback) from a spiral arm that is winding up and dispersing \citep{2016MNRAS.460.2472B}.
This convergence is thought to happen due to a combination of orbit crowding and the gravitational influence of the stellar component of the spiral arm.
The SDW model predicts a global flow through each arm; the dynamic model predicts local flows converging or diverging from each arm. This distinction is why the velocity field of the ISM is a powerful discriminator.

To have a quantity that can be directly compared to a gas velocity field measurement, we have collected ten simulations of spiral structure in Milky Way-like galaxies.
Five of these are SDW simulations and five are dynamic simulations.
These simulations were tuned by their authors to match certain observations of the Milky Way but are not considered to be perfect matches.
Our primary observable, the line-of-sight velocity of interestellar matter as a function of position in the Galaxy (which we call $v_d(x, y)$), was not directly used to tune any of the simulations and so can be considered a prediction.

We construct empirical maps of the Milky Way's $v_d(x, y)$ field and compare them with predictions for $v_d(x, y)$ from the simulations.
The maps are made using a collection of techniques we call ``Kinetic Tomography'' (KT).
In \citet[][henceforth TP17]{Tchernyshyov_2017_kt1}, we developed a method for combining measurements of \HI{} and CO emission in ($\ell,b,v_d$) space with the three-dimensional ($\ell, b, d$) reddening map of \citet{2015ApJ...810...25G} to produce a map of $v_d(\ell, b, d)$.
We will call this method ``gas and dust KT'' (G\&D KT) and the resulting map the G\&D KT map.
In TP17, we validated the G\&D KT map in regions containing very dense gas.
To discriminate between theories of spiral structure, we also need to be sure that the $v_d$ map is correct in more diffuse regions.

The main observational contribution of this work is a second map of the Milky Way ISM velocity field, which we will compare to the G\&D KT map and simulations of spiral structure.
This map is based on ISM absorption lines in spectra of stars with known distances.
An ISM absorption line provides the sightline-integrated distribution of its carrier species with respect to $v_d$.
By taking differences between the optical depth profiles of the ISM along approximately the same sightline but with different terminal distances (i.e., different stellar distances), we can get a measurement of the average $v_d$ of the ISM between the endpoints of those sightlines.
With enough background stars, a more sophisticated version of this procedure can be used to make a continuous map of $v_d$.
The absorption line we use is the 1.527 $\mu$m diffuse interstellar band (DIB) \citep[][henceforth Z+15]{Geballe_2011_IRdibs,Zasowski:2015hi} in stellar spectra from the APOGEE survey \citep[Section~\ref{subsec:dib_in_apogee};][]{Majewski_2017_apogeeoverview}.
APOGEE covers much of the Northern Galactic plane and, because it is a near infrared survey, can obtain high resolution, high signal-to-noise ratio spectra of distant, highly reddened stars.
In the first half of this paper, we analyze DIB absorption in APOGEE spectra to produce a map of the local (i.e., non-integrated) line-of-sight velocity using a procedure we will call DIB KT.

The rest of the paper is organized as follows:
In Section \ref{sec:data}, we describe the datasets we use.
In Section \ref{sec:dibkt}, we explain our map-making procedure, and
in Section \ref{sec:comparison-models}, we list and describe the simulations to which we compare our $v_d$ maps.
In Section \ref{sec:results}, we check the DIB KT map against the G\&D KT map.
In Section \ref{sec:discussion}, we compare the DIB KT and G\&D KT maps with $v_d$ maps from simulations and argue that the Milky Way has dynamic, rather than SDW, spiral structure.
Finally, in Section \ref{sec:conclusion}, we conclude.
Throughout this work, we assume a Sun-Galactic center separation of 8.5 kpc, a Galactic rotation rate of 220~km~s$^{-1}$, and a motion of the Sun relative to the local standard of rest (LSR) of 12 km/s towards the Galactic center, 9 km/s in the direction of Galactic rotation at the Sun, and 7 km/s towards the North Galactic pole.

\begin{figure}
\includegraphics[width=\linewidth]{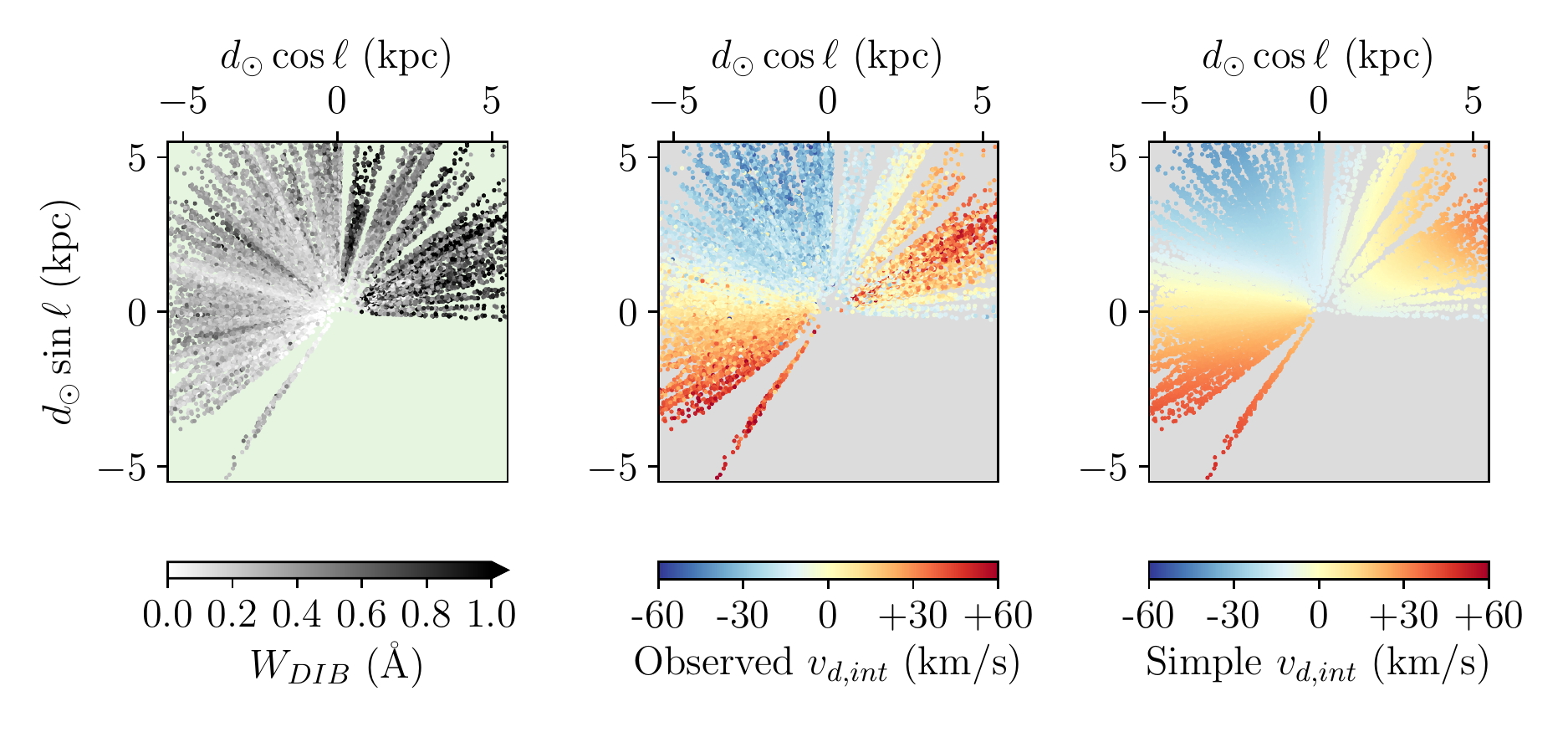}
\label{fig:integrated-quantitites}
\caption{(Left) The equivalent width of the DIB feature towards stars in our subsample of the APOGEE survey. The Galactic center is to the right and the direction of Galactic rotation at the position of the Sun is upwards. (Middle) The first moment $v_{d,int}$ of the DIB feature towards stars in our dataset. The velocity $v_{d,int}$ is the DIB density-weighted average of the (local, un-integrated) velocity $v_d$ along the line of sight. (Right) The $v_{d,int}$ field of a uniform interstellar medium undergoing flat 220 km/s rotation. The measured and computed $v_{d,int}$ are both in the heliocentric rest frame.
}
\end{figure}

\section{Data}
\label{sec:data}
The new map of $v_d(x, y)$ is based on an analysis of absorption by the 1.527~\micron{} DIB in APOGEE spectral residuals.
We describe the APOGEE data and our procedure for selecting and reducing a subsample of the full APOGEE dataset in Section~\ref{subsec:dib_in_apogee}, the characteristics of the 1.527~\micron{} DIB that make it useful for this sort of analysis in Section~\ref{subsec:the-dib}, and our procedure for obtaining distances to stars in our sample in Section~\ref{subsec:stellar-distances}.
In Sections~\ref{subsec:kti-map} and \ref{subsec-reid-hmsfrs}, we describe two prior measurements of the ISM velocity field.
The first of these is the G\&D KT $v_d(\ell, b, d)$ map derived from \HI{} and CO emission and dust reddening derived in TP17 (Section~\ref{subsec:kti-map}).
The second is collection of measurements of the velocities and parallax distances of high mass star formation regions (HMSFRs; Section~\ref{subsec-reid-hmsfrs}).

\subsection{APOGEE Spectra}
\label{subsec:dib_in_apogee}

We measure the 1.527~$\mu$m DIB profiles in spectra from the Apache Point Observatory Galactic Evolution Experiment \citep[APOGEE;][]{Majewski_2017_apogeeoverview}, part of the Sloan Digital Sky Survey III \citep[SDSS-III;][]{Eisenstein_11_sdss3overview} and SDSS-IV \citep{Blanton_2017_sdss4}. This dataset comprises high-resolution $H$-band spectra and derived radial velocities (RVs), stellar parameters, and chemical abundances for $\sim$263,000 stars in the Milky Way (MW) and Local Group, released as part of SDSS Data Release 14 \citep{Abolfathi_2017_SDSSDR14}.  See \citet{Zasowski_2013_apogeetargeting,Zasowski_2017_apogee2targeting} for details of the APOGEE targeting selection, \citet{Nidever_2015_apogeereduction} for a description of the custom data reduction and RV pipeline, \citet{GarciaPerez_2016_aspcap} for details about APOGEE Stellar Parameters and Chemical Abundances Pipeline (ASPCAP), and \citet[][2018 in press]{Holtzman_2015_apogeedata} for details about the data calibration and released data products.

The DIB analysis in Z+15 used the synthetic spectra fits of ASPCAP to remove the stellar absorption lines from APOGEE spectra and isolate the interstellar absorption features (\S\ref{subsec:the-dib}).  In this analysis, we instead adopt stellar spectral models generated by the {\it Cannon} \citep{Ness_2015_Cannon,Casey_2016_cannon2}, which we found to produce cleaner spectral residuals at lower SNR for a wide range of stellar types.

To train our data-driven model, we selected stars with high-quality spectra and reliable stellar parameters, requiring ${\rm SNR} \ge 100$, valid calibrated $T_{\rm eff}$ and $\log{g}$ values, and that the METALS\_BAD, ALPHAFE\_BAD, and STARBAD flags not  be set.
We also required that the stars have Galactic latitudes $|b| \ge 60^\circ$, that the spectra have been observed with the Sloan Foundation 2.5-meter telescope rather than the NMSU 1-meter telescope to ensure consistency with the rest of the sample, and that the stars not have a significant DIB detection in the DIB feature catalog of Z+15.
The $|b|$ limits and quality criteria ensure that our spectral training set does not contain DIB signatures or significant noise that would be carried into the test set's spectral models.  Even though the interstellar and stellar absorption lines are theoretically independent and come from physically distinct sources, some correlation at the pixel level is expected due to the selection-induced correlation between, e.g., DIB strength and stellar temperature; stars need to be more luminous, and hence cooler, to be seen at large distances and behind large amounts of interstellar material.
Thus, eliminating as much DIB contamination as possible from the training set is critical to creating the cleanest residual profiles for this analysis.

The criteria above yielded a training set of 8900 stellar spectra.  We trained a four-label {\it Cannon} model ($T_{\rm eff}$, $\log{g}$, [M/H], and [$\alpha$/M]) on these spectra using the ``Annie's Lasso'' version of the {\it Cannon}\footnote{\url{https://github.com/andycasey/AnniesLasso}}.
We then applied this model to all midplane DR14 spectra (with $|b| \le 1^\circ$) and generated best-fit spectra for each sightline, in addition to the computed stellar labels.  The ratio of each observed spectrum with its {\it Cannon}-derived counterpart is the residual spectrum in which we identify DIB absorption.
Of the full $|b| < 1^\circ$ sample, we keep the 17546 stars for which this instance of the Cannon finds non-null values for all four stellar parameters.

\subsection{The 1.527 \micron{} Diffuse Interstellar Band}
\label{subsec:the-dib}
The 1.527 \micron{} DIB was discovered by \citet{Geballe_2011_IRdibs} in heavily-reddened sightlines towards the Galactic Center.  It is the strongest DIB in APOGEE's $1.5-1.7~\mu$m wavelength range.  Z+15 measured this feature along $\sim$60,000 APOGEE sightlines throughout the Northern sky, deriving its empirical rest wavelength ($\lambda_{\rm 0, vac}=1.5274~\mu$m) and the relationship between its equivalent width ($W_{\rm DIB}$) and dust reddening. \citet{Cox_2014_xshooterDIBs} studied this feature towards a small sample of early-type stars with a range of foreground extinction, and \citet{Elyajouri_2016_apogeetelluricDIBs} extracted and characterized profiles of this DIB from spectra of $\sim$6700 early-type stars in the APOGEE dataset.

Three characteristics of the 1.527~$\mu$m DIB make it particularly useful for an analysis of the ISM velocity field throughout the Galactic plane.
Because its rest wavelength is in the near infrared, where the impact of dust extinction is much weaker than at optical wavelengths \citep[$A(H)/A(V) \approx 0.17$; e.g.,][]{Schlafly_2011_calibSFD}, the DIB can be measured towards stars behind dense molecular gas and/or very long columns of diffuse ISM.
This property enables us to detect the DIB and map the ISM velocity field out to large distances from the Sun.
The equivalent width of the DIB is a near-linear tracer of $A(V)$ at interstellar extinctions typical of even the dusty inner Galactic disk and bulge (Z+15).
\citet{Elyajouri_2017_DIBinB68} find that this linear relationship cannot be applied in clouds with volume density $n_H \gtrsim 10^5$~cm$^{-3}$ such as the Barnard~68 Bok globule.
Fortunately, clouds this dense comprise only a small fraction of the volume of the interstellar medium, which permits us take advantage of the more typical linear behavior for the large-scale mapping performed here.
Finally, the intrinsic profile of the 1.527~$\mu$m DIB is symmetric and consistent.
If the band were asymmetric, this asymmetry would need to be modeled and the degree of asymmetry would be degenerate with the actual value of the velocity field.
This degeneracy is not present when the DIB is symmetric.

To demonstrate that the 1.527$\mu$m DIB contains meaningful and non-trivial information about the ISM velocity field, we show the equivalent width and first moment of the DIB feature towards all of the stars in our sample in Figure \ref{fig:integrated-quantitites}.
For comparison, we also show the integrated velocity field one would expect from a flat 220 km/s rotation curve and a uniform-density ISM.
There are clear differences between the measured and expected sightline-integrated velocities.

\subsection{Stellar distances}
\label{subsec:stellar-distances}

Our analysis requires an estimate of the distance to each star in the sample.
We use a combination of spectrophotometric distances from the APOGEE DR14 red clump catalog \citep{Bovy_2014_APOGEE_RC_catalog} and the DR14 distance Value Added Catalogs \citep[and Holtzman et al, in prep]{Santiago_2016_apogeedistances,Wang_2016_APOGEEdistances} and parallaxes from {\it Gaia} DR2 \citep{GaiaCollab_2016_gaia,GaiaCollab_2018_gaiaDR2}. We collected parallaxes for all APOGEE sources in our sample using the Gaia DR2-2MASS precomputed crossmatch table \citep{Salgado_2017_gaiaarchive,Marrese_2018_gaiaXmatch}.

We combined spectrophotometric distances with parallaxes using a simplified version of the approach described in \citet{McMillan_2018_tgas+rave}. The spectrophotometric distance $d_{sp}$ and its (assumed Gaussian) uncertainty $\sigma_{sp}$ can be thought of as providing a prior for the ``true'' distance $d$ and the parallax $\varpi$, and its uncertainty $\sigma_\varpi$ can be thought of as providing a likelihood for $\varpi$ given $d$.
The posterior probability distribution for $d$ is then given by the expression
\begin{equation}
\label{eqn:single-distance-model}
p(d \vert d_{sp}, \sigma_{sp}, \varpi, \sigma_\varpi) \propto
\mathcal{N}(1/d; \varpi + \delta, \sigma_\varpi^2) \times
\mathcal{N}(d; d_{sp}, \sigma_{sp}^2),
\end{equation}
where $\delta=0.029$ mas is the parallax zero point offset given in \citet{Luri:2018eu} and $\mathcal{N}(d; d_{sp}, \sigma_{sp}^2)$ is a normal distribution with mean $d_{sp}$ and variance $\sigma_{sp}^2$ evaluated at a point $d$. \citet{McMillan_2018_tgas+rave} include stellar parameters, which are covariant with the spectrophotometric distance, in this expression. As we do not have the necessary probability density functions for any of the distance catalogs we use, we assume $d_{sp}$ and $\sigma_{sp}$ provide a sufficient description of the distance probability density function.

There can be multiplicative shifts between the calibrations of distance estimates from different sources.
To infer the value of these shifts between the Gaia DR2 distances and the \citet[][RC]{Bovy_2014_APOGEE_RC_catalog}, \citet[][BPG]{Santiago_2016_apogeedistances}, \citet[][NAOC]{Wang_2016_APOGEEdistances}, and Holtzman et al (in prep, NMSU) spectrophotometric distances, we modify Equation~\ref{eqn:single-distance-model} to include a shift term $a\equiv d/d_\varpi$ and an excess variance term $\sigma_{ext}$:
\begin{equation}
\label{eqn:shifted-distance-model}
p(d_i \vert d_{sp,i}, \sigma_{sp,i}, \varpi_i, \sigma_{\varpi,i}, a, \sigma_{ext}) \propto
\mathcal{N}(1/d_i; \varpi_i + \delta, \sigma_{\varpi,i}^2) \times
\mathcal{N}(d_i; a d_{sp,i}, \sigma_{sp,i}^2 + \sigma_{ext}^2).
\end{equation}
The shift and excess variance terms are assumed to be the same for all spectrophotometric distances determined using a given method.
The per-star variables are given a subscript $i$ to indicate that they vary from star to star.

To avoid complications from unmodeled systematics, we only use stars with parallax signal-to-noise ratios $\varpi / \sigma_\varpi > 25$ and with nominal parallax and spectrophotometric distances that are less than 2 kpc.
We evaluate Equation \ref{eqn:shifted-distance-model} for each star in this subset from a given method over a grid with $d_i$ ranging from 0 to 10 kpc in steps of 0.1 kpc, $a$ ranging from 0.8 to 1.2 in steps of 0.002, and $\sigma_{ext}$ ranging from 0 to 0.5 kpc in steps of 0.01 kpc.
We then integrate over each $d_i$ and over $\sigma_{ext}$ to get a probability density function for $a$.

\begin{figure}
\includegraphics{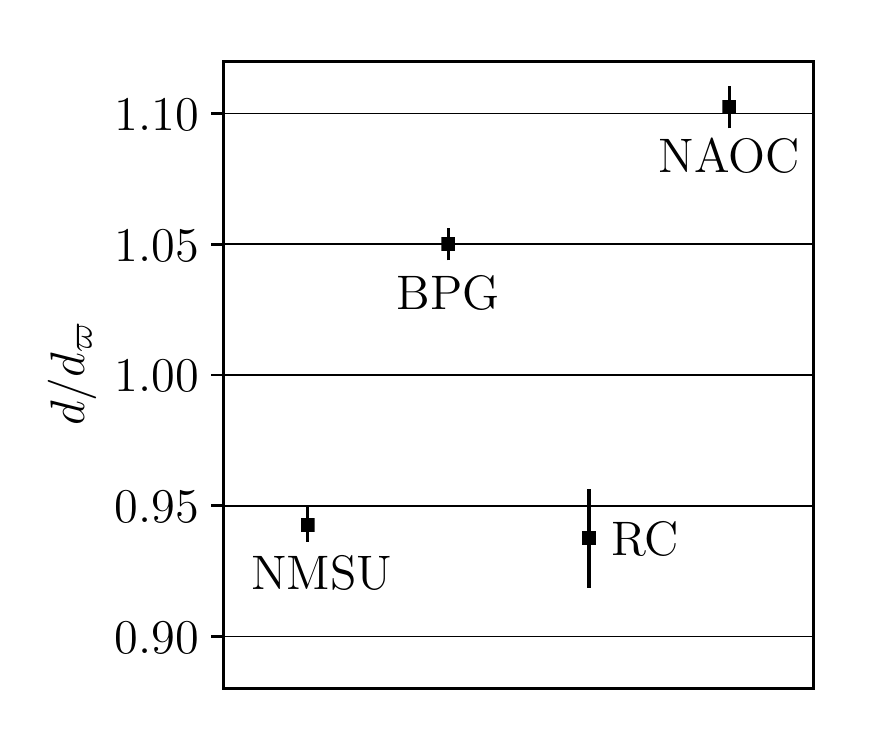}
\label{fig:distance-scale-factors}
\caption{Conversion factors between Gaia DR2 distances (with parallax zero point correction) and four spectrophotometric distance estimation methods from the APOGEE DR14 Value Added Catalogs. The procedure for estimating this conversion factor is described in Section \ref{subsec:stellar-distances}}.
\end{figure}

The expected value and standard deviation of $a$ for each method is shown in Figure \ref{fig:distance-scale-factors}.
The RC shift has a greater uncertainty than the shifts for the other methods because few stars in the RC catalog are within 2 kpc of the Sun.
Because they are mutually consistent, we use the NMSU and RC distances.
We combine the NMSU and RC catalogs with the DR2 parallaxes using the following procedure:
\begin{itemize}
\item If a star has an RC distance and a parallax, shift the RC distance using $a_{RC}$ and combine the result with the parallax using Equation \ref{eqn:single-distance-model}. (3277 stars)
\item If a star has an RC distance but no parallax, shift the RC distance using $a_{RC}$. (53 stars)
\item If a star has an NMSU distance and a parallax but no RC distance, shift the NMSU distance using $a_{NMSU}$ and combine the result with the parallax using Equation \ref{eqn:single-distance-model}. (13226 stars)
\item If a star has an NMSU distance but no parallax or RC distance, shift the NMSU distance using $a_{NMSU}$. (990 stars)
\end{itemize}
The shifts used are the expected values shown as points in Figure \ref{fig:distance-scale-factors}.

\subsection{The Tchernyshyov et al. $v_d(\ell, b, d)$ map}\label{subsec:kti-map}
In TP17, we presented a method for mapping the ISM velocity field using measurements of dust reddening and CO and \HI{} emission. This method was based on matching the amount of dust in voxels of a 3D dust map \citep{2015ApJ...810...25G} with velocity components in a 3D gas emission cube under certain restrictions on the shape of the resulting velocity field. We will be using the ISM velocity map produced in that work (the gas and dust, or G\&D, KT map) to check the DIB-derived velocity map. The two maps were produced using entirely different datasets and methods and should therefore have different systematics, meaning that features of the velocity field that are the same in both maps are most likely real.

The G\&D KT map is 3D and in spherical coordinates while the DIB KT is 2D and in Cartesian coordinates.
We project and regrid the G\&D KT map onto the DIB KT coordinate frame using ISM density-weighted averaging.

\subsection{The Reid et al. high mass star formation regions}\label{subsec-reid-hmsfrs}
R+14 published complete 6D phase space information -- on-sky positions, parallax distances, line-of-sight velocities, and proper motions -- for 103 high mass star forming regions (HMSFRs). In TP17, we used these HMSFRs to test the accuracy of the G\&D KT map. In this work, we will use them to delineate the locus of dense gas and enhanced star formation in the Perseus and Local spiral arms.

\section{Kinetic Tomography with the 1.5 \micron{} DIB}
\label{sec:dibkt}

We generate a map of $v\rmsub{d}(x, y)$ by modeling differences between the DIB spectra (the residual spectra of Section~\ref{subsec:dib_in_apogee}) of pairs of stars with small angular and line-of-sight distance separations.
These DIB difference spectra are an approximation to the derivative of the DIB carrier optical depth distribution $\tau\rmsub{DIB} (\ell, b, d, v_d)$ with respect to distance, $\frac{\dd \tau\rmsub{DIB}} {\dd d}$. $v\rmsub{d}(\ell, b, d)$ is the first moment of $\frac{\dd \tau\rmsub{DIB}} {\dd d}$.
The key assumption of our method is that $\frac{\dd \tau\rmsub{DIB}} {\dd d}$ has a fixed profile, though not necessarily a fixed amplitude, over small regions of space.
We assume this profile is a Gaussian function in $v_d$.
Our method consists of two parts: assigning pairs of sightlines to pixels in the Galactic $(x, y)$ plane (Section~\ref{subsec:pixel-assignment}) and inferring each pixel's $\frac{\dd \tau\rmsub{DIB}} {\dd d}$ profile (Sections \ref{subsec:likelihood-function} through \ref{subsec:outlier-rejection}).

\subsection{Pixel assignment}
\label{subsec:pixel-assignment}
Assigning pairs of sightlines to pixels also consists of two parts.
First, we find pairs of stars whose angular separation $\delta$ is less than some threshold $\delta\rmsub{max}$.
The smaller a pair's $\delta$, the more likely it is that that pair's DIB difference spectrum consists mostly of DIB absorption from ISM between the stars rather than from differences in the spatial distribution of the foreground ISM.
The smaller the adopted $\delta\rmsub{max}$ is, the purer the DIB difference spectra will be; the greater $\delta\rmsub{max}$ is, the more accepted pairs there will be.

We then divide the Galactic plane into square pixels in $x-y$ and assign the pairs with $\delta \le \delta_{\rm max}$ to at most one pixel.
A pair of stars is assigned to a given pixel if the probability $p(>0.5\times {\rm path} \in {\rm pixel}) \equiv p_{\rm pair,\,pix}$ is greater than a threshold $p\rmsub{min}$.
To compute this probability for a given pair of stars, we generate realizations of the path between the stars according to the stars' distances and distance uncertainties.
For each realization, we compute the fraction of the path that falls within each pixel.
We then combine the realizations by computing $p_{\rm pair,\,pix}$, the fraction of the realizations in which more than half of the path falls within each pixel.
If there is a pixel for which $p_{\rm pair,\,pix} > p_{\rm min}$, the stellar pair is assigned to that pixel.
If there is no such pixel, the pair is not used.

This pixel assignment procedure depends on three parameters: the maximum angular separation $\delta\rmsub{max}$, the pixel sidelength  $\Delta x \equiv \Delta y$, and the minimum probability $p\rmsub{min}$.
For our final $v_d(x, y)$ map, we set $\delta\rmsub{max}=0.3^\circ$, $\Delta x = \sfrac{10}{21}$ kpc, and $p\rmsub{min}=0.5$.
These parameter choices represent a tradeoff between the purity of the DIB difference spectra and pixel assignments and the number of DIB difference spectra available for analysis.
As we will show in Section \ref{sec:results}, varying these parameters over a reasonable range does not significantly change the resulting $v_d(x, y)$ map.

\subsection{Likelihood function}
\label{subsec:likelihood-function}
Once the stellar pairs have been assigned, we use Bayesian inference to determine $v_d$ for each pixel. This inference involves four separate models:
\begin{itemize}
  \item A pixel-level model (Section \ref{subsec:pixel-level-model}), which is used to determine $v_d$ for each map pixel.
  \item A model that uses (non-difference) DIB spectra towards nearby stars to determine the distribution of DIB widths (Section \ref{subsec:sigmav-prior}). The width of the DIB is not constant but some widths are more common than others. Including this information as a prior improves the precision and accuracy of the pixel-level solution.
  \item A model that uses all pairs of sightlines that have been assigned to any pixel to determine the distribution of DIB difference spectrum uncertainties (Section \ref{subsec:sigmay-prior}). As with the DIB width, the noise level of the spectra is not constant but also not uniformly distributed.
  \item A model for determining whether a DIB difference spectrum is an outlier (Section \ref{subsec:outlier-rejection}). Some spectra contain extraneous features, such as imperfectly modeled stellar lines, which degrade the quality of the $v_d$ modeling. Removing these spectra increases the precision and accuracy of the pixel-level solution.
\end{itemize}
In principle, these models could be combined into a large single model.
To simplify computation, we have kept them separate.

All four models are based on the same likelihood function.
Given a pair of DIB spectra ${\bf f}_1({\bf v})$ and ${\bf f}_2({\bf v})$, we assume the corresponding DIB difference spectrum ${\bf y} \equiv {\bf f}_1({\bf v}) - {\bf f}_2({\bf v})$ can be described as the sum of a Gaussian function $a \times f(v; v_d, \sigma_v^2)$ with amplitude $a$, center $v_d$, and standard deviation $\sigma_v$; a constant offset $b$; and independent Gaussian noise with standard deviation $\sigma_y$.
We include the constant term $b$ because the DIB-free region of the DIB spectrum can be slightly offset from zero due to imperfect modeling of the stellar spectrum.
The likelihood function for $\bf y$ given these assumptions is
\begin{equation}
  \label{eqn:single-unmarg-likelihood}
  p({\bf y} \equiv \{y_1, y_2, \ldots, y_N \} \vert a, v_d, \sigma_v, b, \sigma_y) =
  \prod_{i=1}^{N} \mathcal{N}\left(y_i; a \times f(v_i; v_d, \sigma_v^2) + b, \sigma_y^2\right),
\end{equation}
where $\mathcal{N}(y_i; a \times f(v_i; v_d, \sigma_v^2) + b, \sigma_y^2)$ is a normal distribution with mean $a \times f(v_i; v_d, \sigma_v^2) + b$ and standard deviation $\sigma_y$ evaluated at the velocity of the $i$th pixel $v_i$.

\subsection{Pixel-level model}
\label{subsec:pixel-level-model}

In the pixel-level model for $v_d$, we assume that all DIB difference spectra that have been assigned to the same pixel have the same $v_d$ and $\sigma_v$ but that each DIB difference spectrum has its own ampliutde $a$, offset $b$, and noise level $\sigma_y$.
Our priors on $a$ and $b$ are Gaussian distributions with mean 0 and standard deviations $\sigma_a=0.5$ and $\sigma_b=0.01$.
This prior on $a$ is essentially uninformative because the maximum amplitude of a DIB difference spectrum in our sample is approximately 0.25.
The prior on $b$ is based on a by-eye estimate of the typical continuum offset.
For most $\bf y$, changing $\sigma_b$ from 0.01 to 1 does not appreciably change the posterior probability distribution for $v_d$.
Our prior on the uncertainty of $\bf y$ is parametrized in terms of the precision $\tau_y \equiv 1 / \sigma_y^2$ rather than the standard deviation $\sigma_y$.
The prior on $\tau_y$ is a truncated gamma distribution with range $1$ to $20000$,  shape parameter $\alpha=1.47$, and rate parameter $\beta=0.95$.
We use a truncated, rather than full, gamma distribution because we marginalize over $\tau_y$ using numerical integration on a fixed (and finite) grid.
The shape and rate parameters of the prior on $\tau_y$ are set in Section \ref{subsec:sigmay-prior}.

The likelihood of the $M$ DIB difference spectra ${\bf y}_1, {\bf y}_2, \ldots, {\bf y}_M$ that have been assigned to a single pixel, given the pixel-level parameters, is
\begin{align}
  \label{eqn:pixel-likelihood}
  \begin{split}
  p(&{\bf y}_1, {\bf y}_2, \ldots, {\bf y}_M \vert v_d, \sigma_v, \sigma_a, \sigma_b, \alpha, \beta) =
  \prod_{j=1}^M p({\bf y}_j \vert v_d, \sigma_v, \sigma_a, \sigma_b, \alpha, \beta)\\
  &= \prod_{j=1}^M \int_{1}^{2/0.01^2} \int_{-\infty}^{\infty} \int_{-\infty}^{\infty}
  p({\bf y}_j \vert a_j, b_j, v_d, \sigma_v, \sigma_{y,j}) p(a_j \vert \sigma_a) p(b_j \vert \sigma_b)
  p(1/\sigma_{y,j}^2 \vert \alpha, \beta) \, \dd a_j \, \dd b_j \, \dd (1/\sigma_{y,j}^2).
\end{split}
\end{align}
The integrals over the amplitudes $a_j$ and offsets $b_j$ have an analytic solution, which we give in Appendix \ref{appendix:analytic-marginalization}.
The integral over $\sigma_{y,j}$ is done numerically.

The prior for $v_d$ is a uniform distribution between $v\rmsub{min}= v_{d,{\rm rot}}(x, y) - 40$ km~s$^{-1}$ and $v\rmsub{max} = v_{d,{\rm rot}}(x, y) + 40$ km~s$^{-1}$, where $v_{d,{\rm rot}}(x, y)$ is the line-of-sight velocity corresponding to rotation according to our fiducial rotation curve at the center of each pixel.
The prior for $\sigma_v$ is a truncated log-normal distribution with range 10 to 50 km~s$^{-1}$, mean parameter $m=3.44$ and standard deviation parameter $s=0.17$.
These parameters are set in Section \ref{subsec:sigmav-prior}.
The posterior probability for $v_d$ in the pixel-level model is then
\begin{align}
  \label{eqn:pixel-level-posterior}
  \begin{split}
  p(v_d \vert {\bf y}_1,& {\bf y}_2, \ldots, {\bf y}_M, \sigma_a, \sigma_b, \alpha, \beta, m, s) \propto \\
  &\int_{10 \text{ km/s}}^{50 \text{ km/s}}
  \left( \prod_{j=1}^M p({\bf y}_j \vert v_d, \sigma_v, \sigma_a, \sigma_b, \alpha, \beta) \right) p(\sigma_v \vert m, s) p(v_d \vert v\rmsub{min}, v\rmsub{max}) \, \dd \sigma_v.
  \end{split}
\end{align}
The integral over $\sigma_v$ is done numerically.

\subsection{Setting the $\sigma_v$ prior}
\label{subsec:sigmav-prior}

We set the parameters of the prior on $\sigma_v$ by modeling the distribution of widths of DIB absorption towards stars within 1 kpc of the Sun.
The pathlength between the Sun and each of these stars is comparable to the typical pathlength between the stars in pairs assigned to a pixel.
The width and amplitude of the total DIB absorption towards these stars should therefore be comparable to the absorption in a DIB difference spectrum $\bf y$.
We set the $\sigma_v$ prior parameters using these (non-difference) DIB spectra, rather than DIB difference spectra, for two reasons.
First, the non-difference spectra will usually have a higher signal-to-noise ratio than the difference spectra.
Second, we know that the DIB absorption towards a nearby star comes from the path between the Sun and that star, while the absorption in a DIB difference spectrum may, despite our assumptions, be partially corrupted by a mismatch between the foreground DIB absorption towards the corresponding pair of stars.

The model for these nearby stars uses the same likelihood function as the pixel-level model (Section \ref{subsec:pixel-level-model}) but has some different priors and a different model structure.
The prior on the Gaussian amplitude $a$ has the same standard deviation as in the pixel-level model but is now a half-normal distribution rather than a normal distribution because the sign of the DIB absorption should be positive.
The prior on the offset $b$ is still a normal distribution but has a smaller standard deviation, $\sigma_b = 0.01 / \sqrt{2}$, because the standard deviation on the baseline of a single DIB spectrum should be a factor of $\sqrt{2}$ smaller than the standard deviation on the baseline of the difference of two DIB spectra.
The prior on $\sigma_y$ is a gamma distribution over $1/\sigma_y^2$ with parameters $\alpha'$ and $\beta'$, which are allowed to vary.
We do not use the same $\sigma_y$ prior as for the pixel-level model because the noise properties of single DIB spectra towards nearby stars will be different from the noise properties of DIB difference spectra of stars from a wider range of distances.
The prior on $v_d$ is defined as before, $v_{d,\rm rot} \pm 40\text{ km~s$^{-1}$}$, but $v_{d,\rm rot}$ is evaluated at the location of the star rather than at the center of a pixel.
The prior on $\sigma_v$ is a log-normal distribution with mean parameter $m$ and standard deviation parameter $s$, which are allowed to vary.
Unlike in the pixel-level model, each sightline has its own $v_d$ and $\sigma_v$ which are marginalized over separately for each star rather than in a tied way for a collection of stars.

The posterior probability distribution for $m$ and $s$ in this model is
\begin{align}
  \label{eqn:sigmav-posterior}
  \begin{split}
  p(&m, s \vert {\bf y}_1, {\bf y}_2, \ldots, {\bf y}_M, \sigma_a, \sigma_b) \propto p(m, s) \times \\
  &
  \int_{\alpha\rmsub{min}}^{\alpha\rmsub{max}}
  \int_{\beta\rmsub{min}}^{\beta\rmsub{max}}
  \prod_{j=1}^M \left(
  \int_{\sigma_{v,\rm min}}^{\sigma_{v,\rm max}}
  \int_{v\rmsub{min}}^{v\rmsub{max}}
  p({\bf y}_j \vert v_d, \sigma_v, \sigma_a, \sigma_b, \alpha, \beta)  p(\sigma_v \vert m, s) p(v_d) \, \dd v_d \, \dd \sigma_v \right) p(\alpha, \beta) \dd \beta \, \dd \alpha.
  \end{split}
\end{align}
where the ${\bf y}_j$ are now just DIB spectra rather than DIB difference spectra.
Priors on $m$, $s$, $\alpha$, and $\beta$ are uniform distributions between $-\infty$ and $\ln 70$, 0 and 10, 0 and $+\infty$, and 0 and $+\infty$.
The integrals over $v_d$ and $\sigma_v$ are done numerically on a fixed grid.
We draw samples from the distribution over $m$ and $s$ and integrate over $\alpha$ and $\beta$ using the \texttt{emcee} implementation of the affine-invariant ensemble sampler for Markov chain Monte Carlo \citep{2013PASP..125..306F}.
The means and standard deviations of the distributions of $m$ and $s$ are $3.44 \pm 0.02$ and $0.17 \pm 0.02$.

We have repeated this procedure using all stars within successively smaller volumes around the Sun: $d<$ 0.9, 0.8, 0.7, 0.6, and 0.5 kpc.
For all cases except 0.5 kpc, the mean values of $m$ and $s$ are the same as that of $d<1$~kpc to within the standard error on the mean of each parameter.
For the 0.5 kpc case, there are not enough observations to provide much of a constraint on $m$ or $s$.
The standard deviations of the $m$ and $s$ distributions are over 50 and over 5, respectively.
We use the mean values of the 1 kpc $m$ and $s$ distributions for the prior on $\sigma_v$ in the pixel-level model (Section~\ref{subsec:pixel-level-model}).

\subsection{Setting the $\sigma_y$ prior}
\label{subsec:sigmay-prior}

We apply a similar procedure to all of the DIB difference spectra to set the parameters of the prior on $\sigma_y$.
We assume that anything that is not captured by our usual model for a DIB difference spectrum, a single Gaussian superimposed on a constant baseline, is noise.
As in the procedure for determining the prior on $\sigma_v$, we assume each DIB difference spectrum has its own $v_d$ and $\sigma_v$.
Our prior on each $v_d$ is once again $v_{d,\rm rot} \pm 40\text{ km~s$^{-1}$}$, where we evaluate $v_{d,\rm rot}$ at the center of the $0.5 \times 0.5$~kpc pixel to which the stellar pair in question has been assigned.
The priors on $a$, $b$, and $\sigma_v$ are the same as in the pixel-level model (Section~\ref{subsec:pixel-level-model}).

With these priors and model structure, the posterior probability distribution for $\alpha$ and $\beta$ is
\begin{align}
  \label{eqn:sigmay-posterior}
  \begin{split}
  p(m,& s \vert {\bf y}_1, {\bf y}_2, \ldots, {\bf y}_M, \sigma_a, \sigma_b) \propto p(\alpha, \beta) \times \\
  &
  \prod_{j=1}^M \left(
  \int_{\sigma_{v,\rm min}}^{\sigma_{v,\rm max}}
  \int_{v\rmsub{min}}^{v\rmsub{max}}
  p({\bf y}_j \vert v_d, \sigma_v, \sigma_a, \sigma_b, \alpha, \beta)  p(\sigma_v \vert m, s) p(v_d) \, \dd v_d \, \dd \sigma_v \right).
  \end{split}
\end{align}
The priors on $\alpha$ and $\beta$ are uniform distributions between 0 and $+\infty$.
The marginalization over $v_d$ and $\sigma_v$ is done numerically on a fixed grid.
We draw samples from the posterior probability distribution for $\alpha$ and $\beta$ once again using the affine-invariant ensemble sampler implemented in \texttt{emcee}.
The means and standard deviations of these distributions are $1.47 \pm 0.04$ and $0.95 \pm 0.03$.
We use these mean values as the parameters of the $\sigma_y$ prior in the pixel-level model (Section \ref{subsec:pixel-level-model}).

\subsection{Deciding if a difference spectrum is an outlier}
\label{subsec:outlier-rejection}

\begin{figure}
  \includegraphics{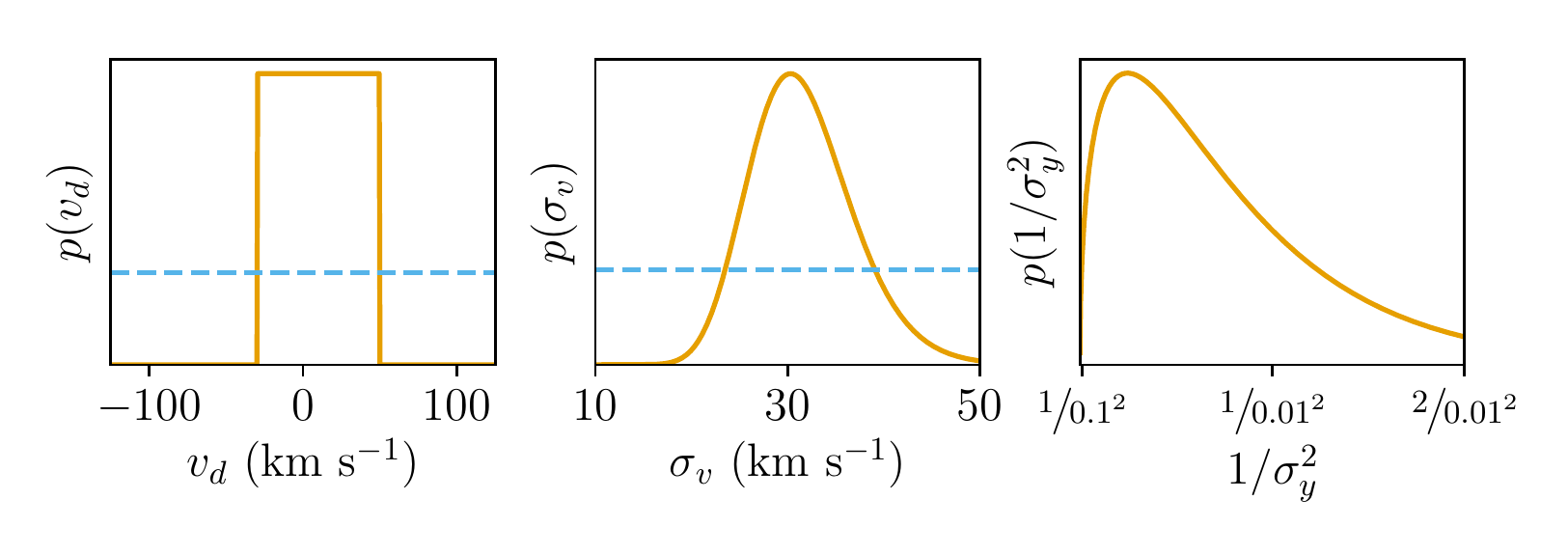}
  \caption{Priors on $v_d$ (left), $\sigma_v$ (center), and $\sfrac{1}{\sigma_v^2}$ (right). Priors used to make the $v_d(x, y)$ map are shown in solid orange. For $v_d$ and $\sigma_v$, flat priors used for deciding whether a given DIB difference spectrum is an outlier (Section~\ref{subsec:outlier-rejection}) are shown in dashed blue. The center of the $v_d$ prior varies depending on location in the Galaxy, but the width remains constant.}
  \label{fig:priors}
\end{figure}

\begin{figure}
  \includegraphics{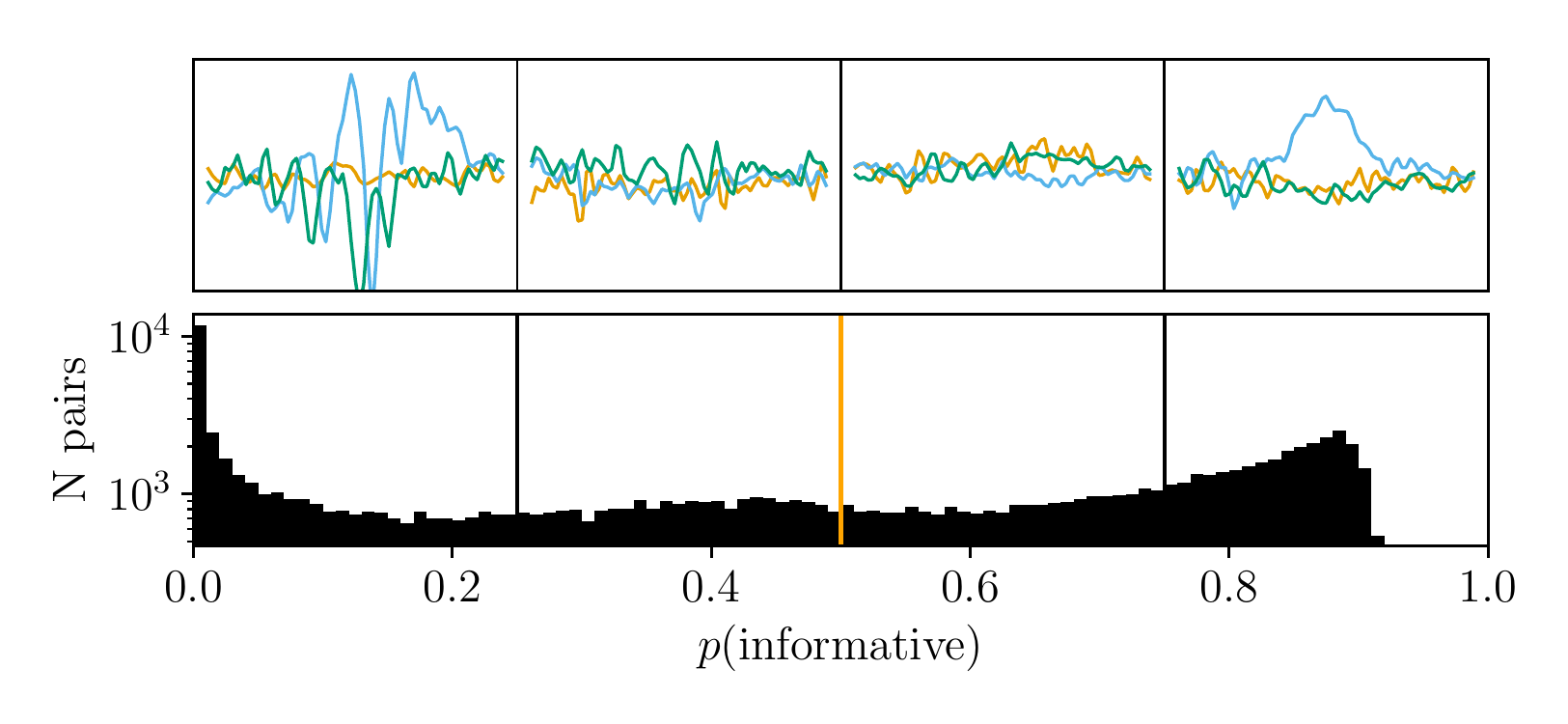}
  \caption{Bottom panel: distribution of $p(\text{informative})$. Top panels, from left to right: examples of DIB difference spectra with $0 \leq p(\text{informative}) < 1/4$, $1/4 \leq p(\text{informative}) < 1/2$, $1/2 \leq p(\text{informative}) < 3/4$, and $3/4 \leq p(\text{informative}) \leq 1$.}
  \label{fig:outlier-demonstration}
\end{figure}

Many of the DIB spectra contain spurious absorption and emission features.
These features are the result of imperfect modeling of stellar, telluric, and sky emission lines but are mistaken to be (possibly quite strong) DIB absorption by our simple model.
Including DIB difference spectra with these sorts of features in pixel-level inference without expanding the model to include them will degrade the quality of the $v_d(x, y)$ map.
While we do not have a model for these features, we do have a way of identifying these contaminated DIB difference spectra.

Stellar, telluric, and sky lines tend to be narrower than DIB absorption and, unlike DIB absorption, can be centered at unphysical velocities in the DIB velocity frame.
The Gaussian parameters that best describe these contaminants are disfavored (in the case of $\sigma_v$) or outright excluded (in the case of $v_d$) by our informative priors.
A contaminated spectrum should therefore be better described by a Gaussian + baseline model with a flat prior on $\sigma_v$ and a broad, flat prior on $v_d$ than by a model with our informative priors.
We show these informative priors on $v_d$, $\sigma_v$, and $\sigma_y$ and flat priors over the same range on $v_d$ and $\sigma_v$ in Figure \ref{fig:priors}.

To determine whether a given DIB difference spectrum is better described by an model with informative priors or flat priors on the Gaussian parameters, we marginalize over all parameters with the two sets of priors to get two marginal likelihoods, $p({\bf y})\rmsub{inf}$ and $p({\bf y})\rmsub{flat}$, for each model.
The probability that the model with informative priors is a better description of the DIB difference spectrum is then
\begin{equation}
  \label{eqn:p-informative}
  p(\text{informative}) = \frac{p({\bf y})\rmsub{inf}}{ p({\bf y})\rmsub{inf} + p({\bf y})\rmsub{flat}},
\end{equation}
where we have implicitly assumed that the informative prior and flat prior models are equally likely a priori.
We show the distribution of $p(\text{informative})$ among all DIB difference spectra that were assigned to any pixel in Figure \ref{fig:outlier-demonstration} along with some example DIB difference spectra from four different $p(\text{informative})$ ranges.
We assume all DIB difference spectra with $p(\text{informative}) > 0.5$ are uncontaminated.
These are the spectra we use to derive the $v_d(x, y)$ map (Section~\ref{sec:results}) using the pixel-level model described in Section~\ref{subsec:pixel-level-model}.

\section{Comparison simulations}
\label{sec:comparison-models}
To help interpret the G\&D KT and DIB KT $v_d(x, y)$ maps, we compare them with $v_d(x, y)$ maps derived from simulations of gas flow in disk galaxies with spiral structure.
These simulations fall into two broad classes --- those with a fixed background potential, corresponding to the SDW model, and those with a dynamically evolving background potential, corresponding to the dynamic model.

Both classes of simulation follow the flow of ISM in a gravitational potential that is set mostly by dark matter and stars.
In the first class, the evolution of the stellar distribution is described analytically.
For example, the stellar distribution can be described as a linear spiral density wave or as a stable bar-induced spiral.
In the second class, the stellar distribution evolves in the same gravitational potential as the gas.
We consider five SDW simulations (Section~\ref{sec:sdw_simulations}) and five dynamic spiral structure simulations (Section~\ref{sec:dynamic_simulations}).

From each simulation, we compute a flat rotation-subtracted $v_d(x, y)$ map.
These $v_d(x, y)$ maps are initially computed at the native resolution of each simulation and then degraded to the resolution of the KT maps using surface density-weighted averaging.
We obtain the rotation rate by taking the mean of the (galactocentric) azimuthal velocity of the simulated ISM in an annulus centered on the (simulated) galactic center.
The annulus extends 0.5 kpc inward and outward from the galactocentric radius of the observer.
We use a flat rotation curve with a rotation rate appropriate for the location of the observer to match what we have done for the KT-derived $v_d(x, y)$ maps.
Computing a radially varying rotation curve from the simulated velocity field would be trivial but would erase features from the simulated $v_d(x, y)$ maps that could potentially be present in the KT-derived $v_d(x, y)$ maps.

\subsection{Stationary density wave simulations}
\label{sec:sdw_simulations}
All five of the SDW simulations include spiral~arm-like perturbations, but they differ in the number of arms, the properties of the arms, the presence of a bar, and the properties of the bar.
Four of the simulations come from \citet{Pettitt:2014ep} and one comes from \citet{Li:2016dx}.
For each set of simulations, we describe below the shape of the perturbing potential, the included physics, and the solution method used for evolving the equations of hydrodynamics.

\subsubsection{Pettitt et al. (2014)}
\citet{Pettitt:2014ep} made a suite of simulations in an attempt to reproduce the Galactic CO $\ell-v$ diagram.
Here, we examine the four simulations shown in their Figure 25.
These simulations are available at http://hdl.handle.net/10871/15057; we received useful advice on orienting them from the authors (Pettitt 2017, private communication communication).
The axisymmetric part of the potential consists of bulge, halo, and disc terms with amplitudes tuned to reproduce the \citet{2012PASJ...64...75S} rotation curve.
There are four different perturbing potentials, combining two- and four-armed logarithmic spiral perturbations with bars of two different strengths.
While the general purpose of the work was to reproduce spiral features in the $\ell-v$ diagram, it was found that no one simulation was perfect; these four simulations reproduce some, but not all, of the known $\ell-v$ diagram features.
The simulations should be thought of as Milky Way-like rather than as best fits to the Milky Way data.
We will refer to these simulations as P-SDW1 through P-SDW4.

The simulations are computed in 3D and include compressible, inviscid gas hydrodynamics with an adiabatic equation of state, simplified \molH{} and CO chemistry, and ISM heating and cooling.
They do not include star formation feedback or gas self-gravity.
The equations of motion are solved using smoothed particle hydrodynamics.
ISM state variables such as temperature and chemical composition are tracked and evolved independently for each particle.

\subsubsection{Li et al. 2016}
\citet{Li:2016dx} were aiming to reproduce features from l-v diagrams, in particular inner-Galaxy features such as Bania's Clump 2 and the details of the molecular ring (Scutum-Centaurus arm).
We received a snapshot of the simulation shown in Figure 2 of \citet{Li:2016dx} directly from the first author (Li 2017, private communication).
Their potential is built from a bulge model from \citet{2015MNRAS.448..713P}, a nuclear bulge component, four logarithmic spiral arms based on the star formation-traced arms defined in R+14, and a long bar component.
This simulation was tailored to match the R+14 spiral arms and so can be expected to be a more accurate estimate of the density and velocity field of the Milky Way than any of the \citet{Pettitt:2014ep} simulations.
We will refer to this simulation as L-SDW.

This simulation is computed in 2D and includes compressible, inviscid gas hydrodynamics with an isothermal equation of state.
It does not include star formation, gas self-gravity, chemistry, or heating and/or cooling.
The equations of motion are solved using a finite volume method on a fixed Cartesian grid.

\subsection{Dynamic spiral structure simulations}
\label{sec:dynamic_simulations}
In addition to the five stationary density wave simulations with fixed potentials, we also consider five dynamic spiral structure simulations with live, evolving potentials.
In these simulations, the stellar spiral and bar perturbations form spontaneously from an initially cylindrically symmetric configuration.
These spiral perturbations are not stable.
Over the course of a simulation, a given arm will form, persist for some time, and then dissipate.
The presence of spiral structure, however, is stable: individual arms may form and dissipate, and the number of arms may change, but at any given time, there will be morphologically obvious spiral structure.
Characteristics of this spiral structure such as the average number of arms, their amplitude above the baseline, and their typical lifetime are set by the disc-to-halo mass ratio \citep{1984ApJ...282...61S}.
We describe the initial conditions and included physics in more detail below.
In all cases, the simulations are 3D, the live stellar component is simulated using N-body techniques, and the gas component is simulated using smoothed particle hydrodynamics.

\subsubsection{Kawata et al. 2014}
The \citet{Kawata:2014ho} simulation is meant to have Milky Way-like stellar, gaseous, and dark matter masses.
We received a snapshot of this simulation directly from the first author (Kawata 2017, private communication communication).
This particular snapshot was chosen because it has a spiral arm and bar that roughly line up with the Perseus arm and Galactic bar. It is not meant to reproduce specific features of the Milky Way in detail.
We will refer to this simulation as K-D.

The dark matter halo is assumed to be static and there is no bulge component.
The simulation includes metal enrichment, ISM heating and cooling, self-gravity, density threshold-based star formation, stellar (wind) feedback, and supernova feedback.
This is the only simulation we consider that includes star formation and feedback.
Supernova-driven features appear in the simulation's equivalent of the Perseus arm, though we do not compare features between the simulations and observations at that level of detail.

\subsubsection{Pettitt et al. 2015}
The \citet{Pettitt:2015kx} simulations were made in order to accurately reproduce the Milky Way CO $\ell-v$ diagram, particularly away from the nuclear region.
Runs with seven different initial and static mass distributions were performed.
Among all snapshots of these seven simulations, the four best matches to the observed CO $\ell-v$ diagram were chosen.
These best matches are shown in Figure 10 of \citet{Pettitt:2015kx}.
We received the snapshot files directly from the first author (Pettitt, private communication).
We will refer to these simulations as P-D1 through P-D4.

As stated in \citet{Pettitt:2015kx}, the best matches come from simulations with a static dark matter halo and a live stellar bulge and disc.
The simulations include the same basic H2 and CO chemistry and ISM cooling and heating as the \citet{Pettitt:2014ep} simulations.
They do not include gas self-gravity or star formation.

\section{Results}
\label{sec:results}

\begin{figure}
\includegraphics[width=\linewidth]{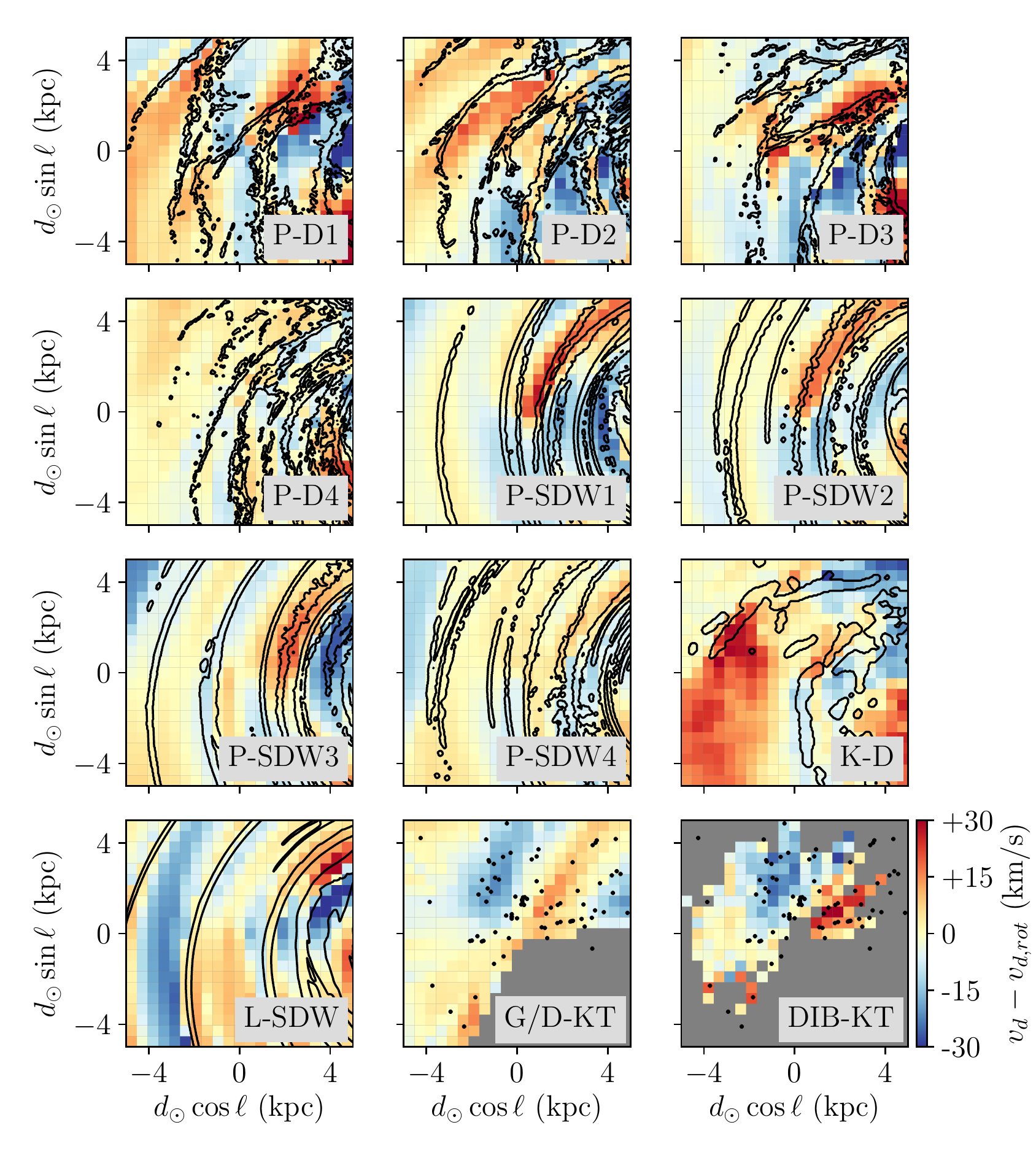}
\label{fig:models-and-KT}
\caption{Flat rotation-subtracted ISM velocity fields measured using Kinetic Tomography (G\&D KT and DIB KT) and predicted by simulations (all other panels). Colors indicate the rotation-subtracted line-of-sight velocities. In the simulation panels, black contours indicate locations where the ISM surface density is in the top decile of surface densities in the simulation domain. In the KT panels, black dots indicate the locations of high mass star formation regions from \citet{2014ApJ...783..130R}. }
\end{figure}

\begin{figure}
  \gridline{\fig{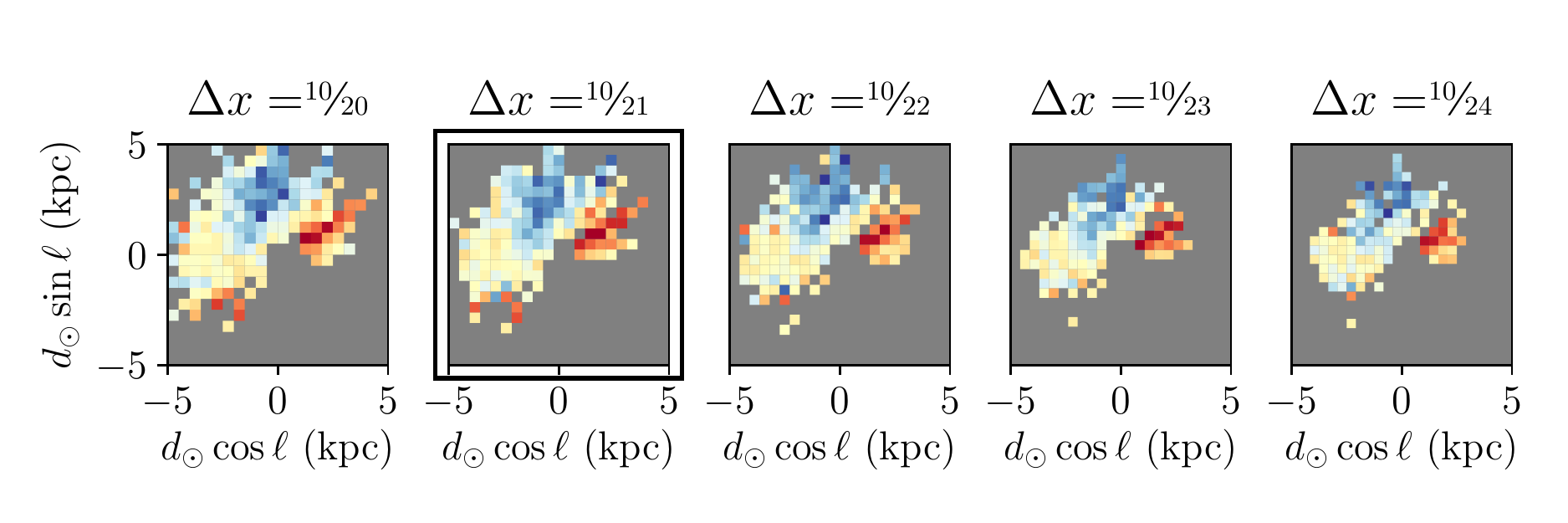}{0.9 \textwidth}{}}
  \gridline{\fig{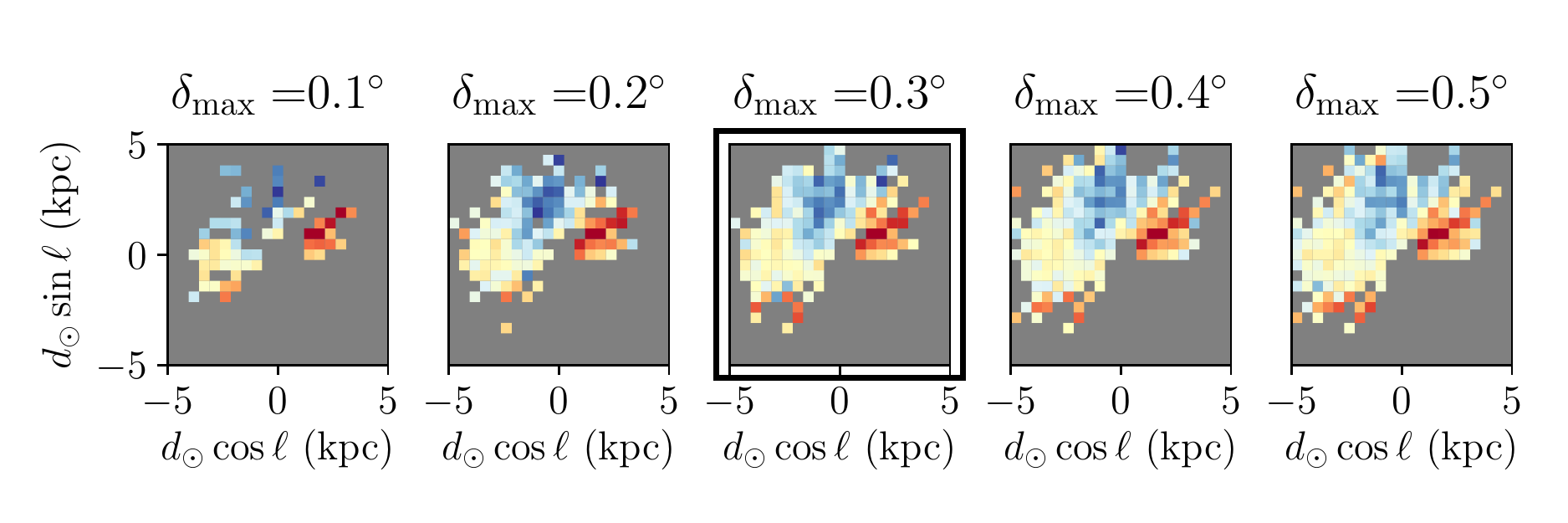}{0.9 \textwidth}{}}
  \gridline{\fig{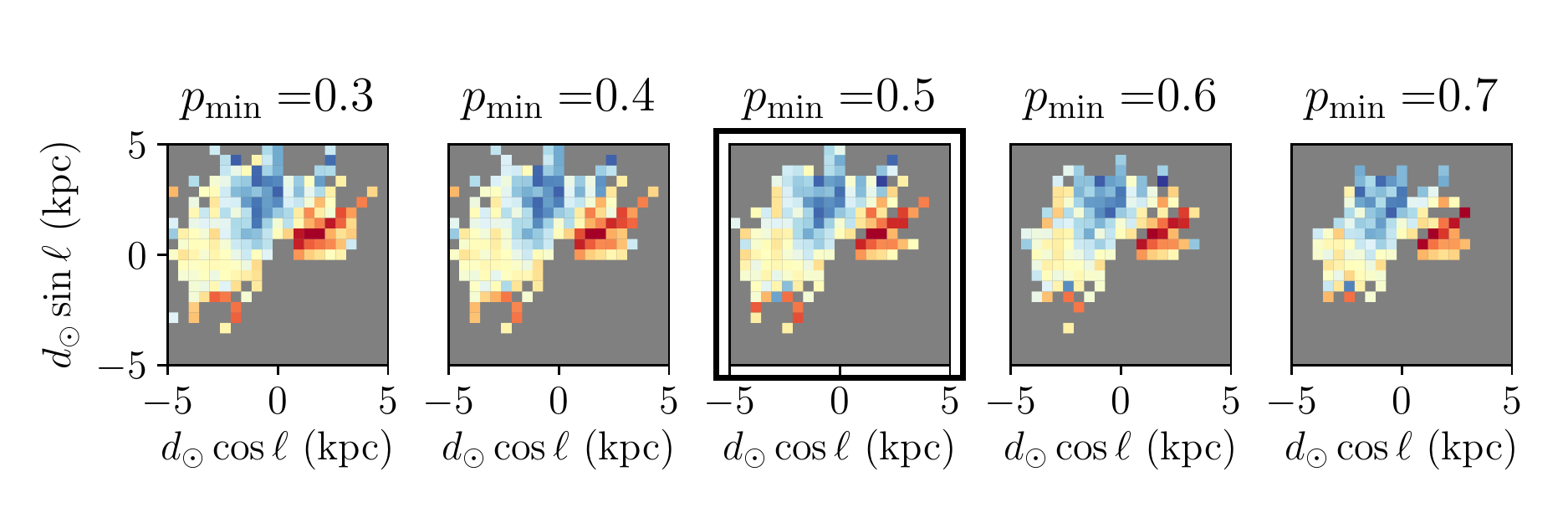}{0.9 \textwidth}{}}
  \caption{The effect of varying one map-making parameter at a time on the DIB KT velocity map. In each row, the map made using the adopted parameters is indicated with a box. {\bf Top:} Map pixel size in kpc. {\bf Middle:} Maximum angular separation between stars in a sightline pair. {\bf Bottom:} Minimum $p\rmsub{pair,\,pix}$ required for a sightline pair to be assigned to a pixel. See Section \ref{subsec:pixel-assignment} for an in-depth description of these parameters and their role in DIB KT.}
  \label{fig:parameter-sequence-KT-maps}
\end{figure}

The primary result of this work, a planar map of $v_d(x, y)$ derived from DIB absorption, is shown in the bottom right panel of Figure~\ref{fig:models-and-KT}.
The DIB KT map and planar version of the G\&D KT map are available at \dataset[10.7910/DVN/UPJM6D]{\doi{10.7910/DVN/UPJM6D}}.
We compare the DIB KT map with the $v_d(x, y)$ maps derived from simulations of spiral structure in Section~\ref{sec:discussion}.
In the current section, we demonstrate that the DIB KT $v_d(x, y)$ map does not strongly depend on the parameters of the map-making method (Section~\ref{sec:param_choices}) and compare the DIB KT and G\&D KT $v_d(x, y)$ maps (Section~\ref{sec:kte_vs_kta}).

\subsection{Parameter choices}
\label{sec:param_choices}
Once star pairs have been assigned to pixels, DIB KT is essentially self-calibrating.
We describe this self-calibration in detail in Sections~\ref{subsec:sigmav-prior}, \ref{subsec:sigmay-prior}, and \ref{subsec:outlier-rejection}.
The procedure for assigning star pairs to pixels depends on three parameters for which we do not have a self-calibration scheme.
These are the pixel sidelength $\Delta x$, the maximum angular separation between stars in a pair $\delta\rmsub{max}$, and the minimum value $p\rmsub{min}$ of $p_{\rm pair,\,pix}$ required for a pair to be assigned to a pixel.

One way to test our choice for $\Delta x$ is to compare the uncertainties on the DIB-derived $v_d(x, y)$ map to the sub-pixel variance of the simulated $v_d(x, y)$ maps at different pixel sizes.
The sub-pixel variance is the variance of the distribution of velocities in a pixel.
As pixel size increases, the uncertainties decrease and the sub-pixel variances increase.
Assigning a single velocity to a pixel is an approximation whose accuracy can be described by the sub-pixel variance.
When the $v_d$ uncertainty is smaller than the sub-pixel variance, the $v_d$ estimate error is dominated by the limitations of the approximation.

\begin{figure}
  \includegraphics{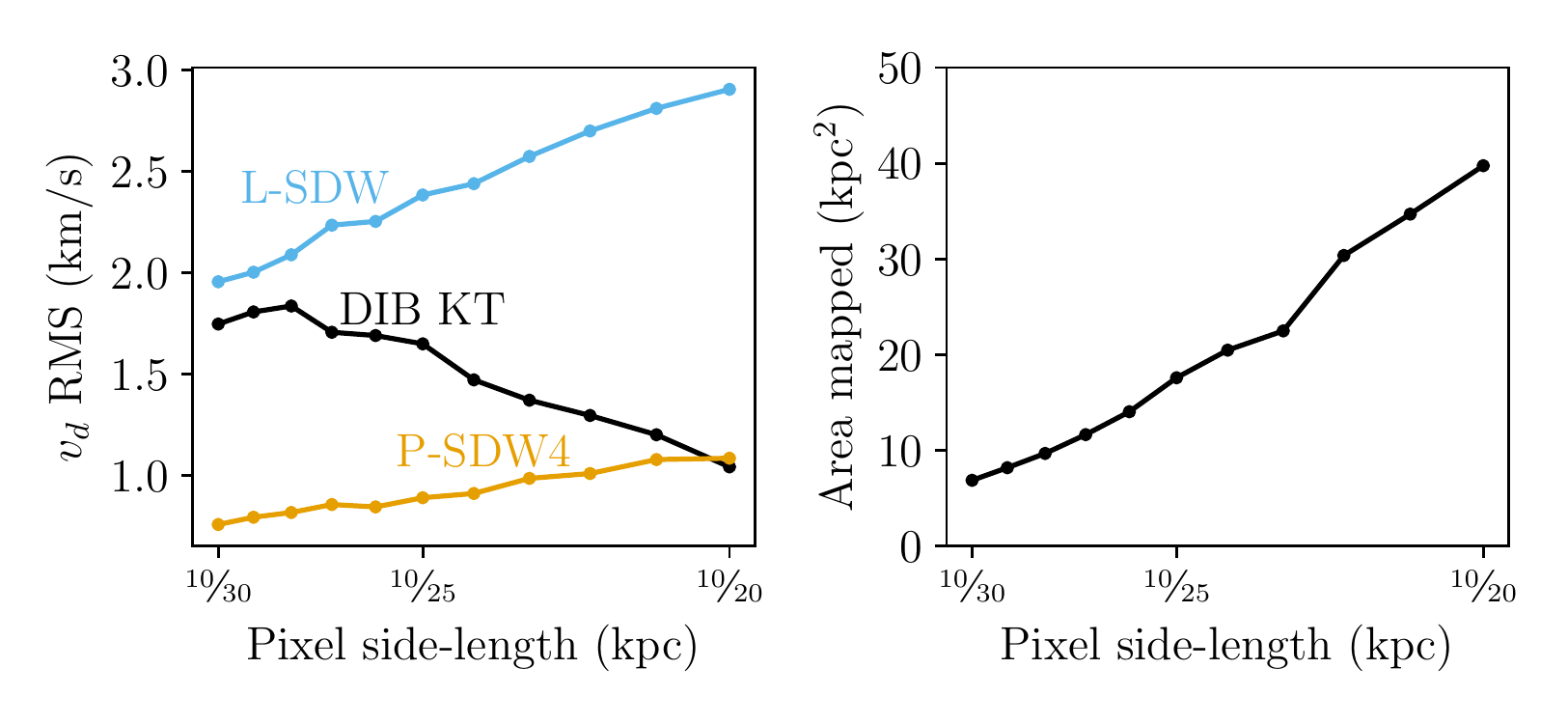}
  \caption{{\bf Left panel:} Median RMS of $v_d$ as a function of pixel size for two simulations and the DIB KT map. For the simulations, the RMS is the standard deviation of the distribution of velocities in each pixel.  For the DIB KT map, the RMS is the standard deviation of the posterior probability distribution of $v_d$ in each pixel. {\bf Right panel:} The area of the DIB KT map as a function of the pixel size.}
  \label{fig:sidelength-trends}
\end{figure}

Figure \ref{fig:sidelength-trends} compares the median uncertainty of the DIB-derived $v_d(x, y)$ map to the median amount of sub-pixel variation in two of the simulations we consider, the \citet{Li:2016dx} spiral density wave (L-SDW) and the fourth \citet{Pettitt:2015kx} spiral density wave (P-SDW4) simulations.
Because the L-SDW simulation has the strongest velocity contrasts among the simulations we consider, it also has the greatest amount of sub-pixel variation.
The P-SDW4 simulation has the weakest velocity contrasts and therefore has the smallest amount of sub-pixel variation.
If we want to be able to confirm or rule out the P-SDW4 simulation, we should pick a pixel size such that the typical uncertainty is not significantly greater than the typical amount of sub-pixel variation.
Conversely, it is not useful to pick a pixel size for which the typical uncertainty is significantly smaller than the typical amount of sub-pixel variation.

A second way to motivate a choice for $\Delta x$ is to compute the area of the $v_d(x, y)$ map as a function of pixel size.
Decreasing the size of the pixels reduces the number of usable star pairs, since it requires pairs of stars to have more precise distances and smaller separations in order to be assigned to a pixel.
This in turn tends to reduce the area over which there are pixels with a sufficient number of pairs of stars to measure $v_d$.
We show how the area depends on the pixel size in Figure \ref{fig:sidelength-trends}.

A $\Delta x$ of \sfrac{10}{22}, \sfrac{10}{21}, or \sfrac{10}{20} kpc would be reasonable according to both of these metrics.
We use a $\Delta x$ of \sfrac{10}{21} kpc.
In Figure \ref{fig:parameter-sequence-KT-maps}, we show $v_d(x, y)$ maps derived assuming five different pixel sizes.
While the area of the maps decreases as the pixel size gets smaller, there is no substantive change in the features of the velocity field.
In the same Figure, we also show how the $v_d(x,y)$ map changes as a result of varying $\delta\rmsub{max}$ and $p\rmsub{min}$.
As with $\Delta x$, there is no substantive change except for a decrease in the area covered by the map.

\subsection{Comparing DIB KT and G\&D KT}
\label{sec:kte_vs_kta}

\begin{figure}
  \includegraphics{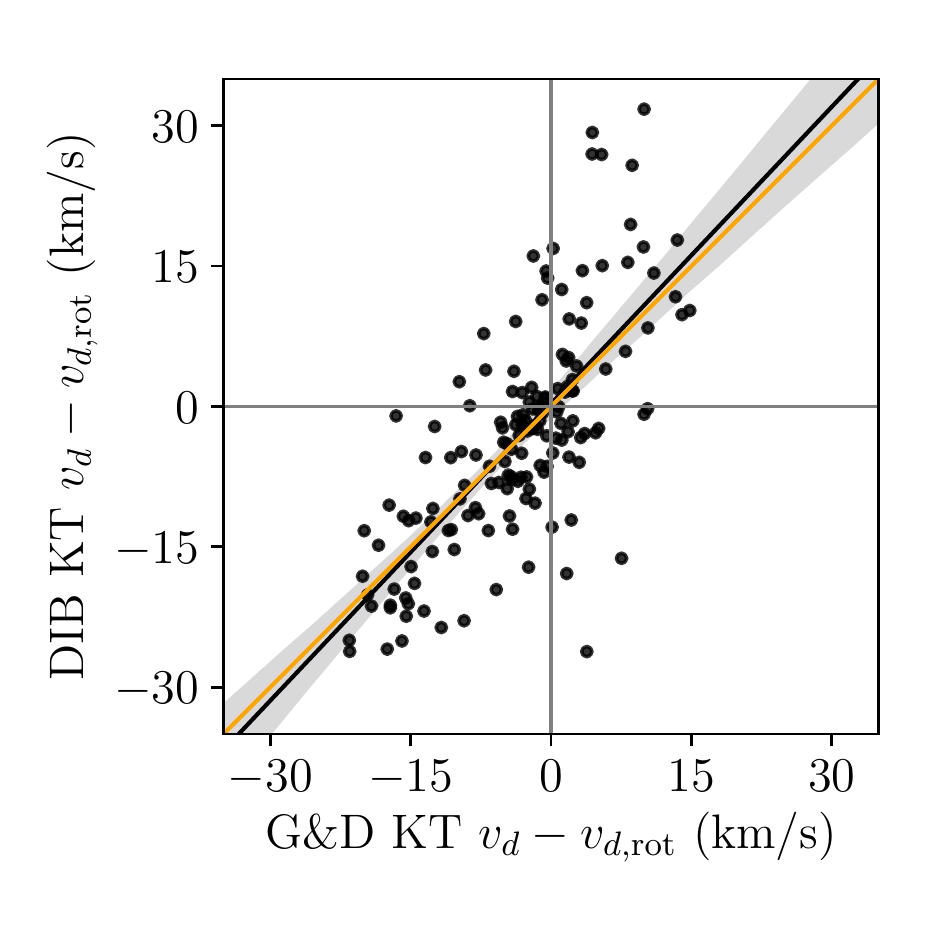}
  \caption{The rotation-subtracted velocities assigned to each pixel by G\&D KT (x-axis) and DIB KT (y-axis). The black line is the best-fit linear relation, the gray filled area is the 95\% confidence interval for the linear relation, and the orange line is a one-to-one relation.}
  \label{fig:gd-vs-dib-kt}
\end{figure}
We can check the DIB KT map by comparing it to the G\&D KT map.
Both KT techniques measure the same quantity, $v_d(x, y)$, but use completely different dataset and techniques.
One simple way to compare the two maps is to directly compare the velocities they assign to pixels where they overlap.
This comparison is shown in Figure \ref{fig:gd-vs-dib-kt}.
While there are points where the maps do not agree by an amount that is greater than what we would estimate is the uncertainty on either measurement, the two maps agree on average --- the slope of the maximum-likelihood linear relation between values from the two maps is consistent with unity.

In addition to being pointwise consistent on average, the maps contain the same qualitative velocity features.
We highlight four major features in the KT maps in Figure \ref{fig:models-and-KT}: a receding (red) region in the 1st quadrant, an appproaching (blue) region in the near part of the 2nd quadrant, a receding or zero velocity region in the far part of the 2nd quadrant, and a receding or zero velocity region in the 3rd quadrant.
The locations and shapes of these features are almost identical in the two maps.
The main differences between shapes of velocity features are in the 1st quadrant, where we expect the G\&D KT map to be less accurate than in the outer galaxy.
The \citet{2015ApJ...810...25G} dust map, which is the source of distance information in G\&D KT, is accurate to greater distances in the lower-average-density outer Galaxy than in the higher-average-density inner Galaxy.
The gas emission cubes, which are the source of velocity information in G\&D KT, are more confused in the inner Galaxy than the outer Galaxy since gas at a given velocity can be located at two very different distances.
These possible issues are explored in more detail in Section 4.2 of TP17.
We conclude that the two KT maps are mutually consistent.

\section{Discussion} \label{sec:discussion}

As can be seen in Figure \ref{fig:models-and-KT}, none of the simulations provide an exact match to our inferred velocity field.
The two simulations which come closest to reproducing the KT velocity field are P-D1 and P-SDW3.
We will argue below that the differences between P-D1 and KT are reasonable and not unexpected in the dynamical spiral structure model while the differences between P-SDW3 and KT are insurmountable in the SDW model.
Furthermore, we will argue that these differences have to be present in any stationary density wave ISM velocity field.
KT is strongly inconsistent with the SDW model but can be consistent with a dynamic spiral structure model in which the Perseus arm near the Sun is in the process of dissipating.

\begin{figure}
  \includegraphics{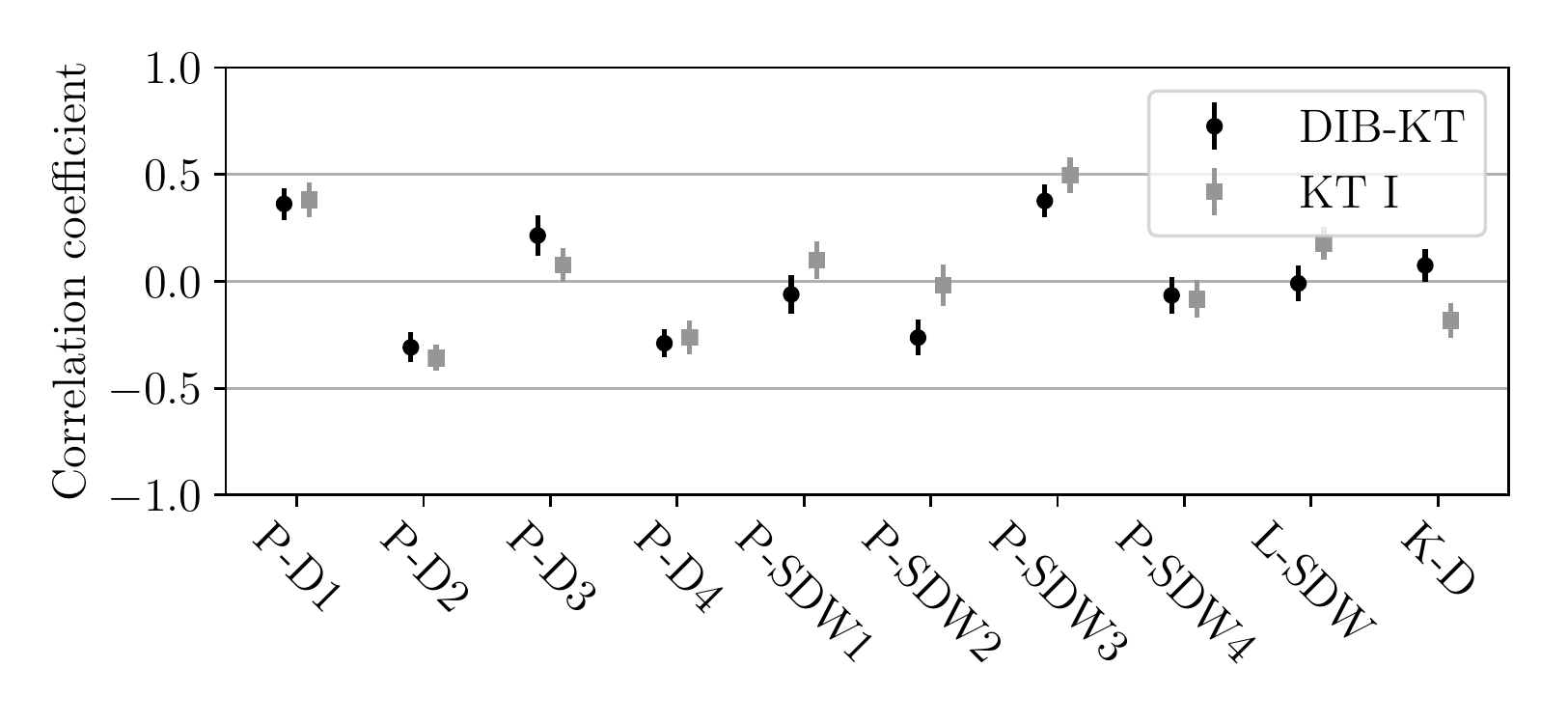}
  \caption{Spearman correlation coefficients computed between the pixel-wise values of $v_d(x, y)$ from G\&D and DIB KT and from spiral structure simulations.}
  \label{fig:KT-model-corr-coeffs}
\end{figure}

The simplest way to quantitatively compare two maps is to compare their values point by point.
We do this point by point comparison by computing the Spearman correlation coefficient between each of the KT-derived $v_d(x, y)$ maps and each of the simulated $v_d(x, y)$ maps.
These coefficients are shown in Figure \ref{fig:KT-model-corr-coeffs}.
An identical pair of maps would have a correlation coefficient of 1.
A map and its opposite, i.e. the result of multiplying each $v_d$ value by -1, would have a correlation coefficient of -1.
The two simulated maps that most closely resemble the DIB and G\&D KT maps according to this metric are P-D1, a dynamic spiral structure simulation from \citet{Pettitt:2015kx}, and P-SDW3, a spiral density wave simulation from \citet{Pettitt:2014ep}.

Point by point comparisons such as this correlation coefficient ignore spatial structure and can be confused by small spatial shifts.
To better understand the similarity and dissimilarity of maps, we look at features of the map that could be driving these coefficients.
Major features in the KT maps include: an approaching region in the near part of second quadrant, a receding region in the first quadrant, and minimal streaming in the third quadrant.
P-D1 and P-SDW3 contain all of these features.
P-D2 and P-D4 have negative correlation coefficients because they have an approaching region in the near part of the second quadrant and an approaching region in the 3rd quadrant.
Simulations such as L-SDW and K-D agree in some parts (near part of the second and first quadrants, respectively) but disagree in others and so get correlation coefficients near 0.
This qualitative comparison is thus consistent with the Spearman correlation analysis.

The P-D1 and P-SDW3 simulations cannot both be correct or close to correct, since they are based on different assumptions about the nature of spiral structure.
To get a better sense of which is correct, we examine broader, less realization-specific predictions for the velocity fields of dynamic spiral structure and SDW spiral structure.
These sorts of predictions are more likely to actually apply to the KT velocity field and can be more constraining than any realization-specific detail.
We also examine the location of dense gas relative to velocity field features.
The dense gas, i.e. the actual gaseous spiral arms, are a natural reference point when trying to compare observations with simulations.
According to these more general predictions and additional information from dense gas, the KT $v_d$ maps favor the dynamic spiral structure simulation over the SDW simulation.

SDW models predict that a gaseous spiral arm is fed by diffuse gas flowing into the arm from one side.
Inside the corotation radius of the spiral pattern, this gas flow should be mostly outward into the trailing edge of the spiral arm.
Outside the corotation radius, the gas flow should be mostly inward into the leading edge of the spiral arm.
The dense gas should be located downstream from this strong radial flow and should have a relatively small peculiar velocity in the Galactocentric radial direction.

These diffuse, mostly radial flows should be consistent over the extent of the spiral arm and should be particularly obvious in regions of the outer Galaxy where $v_d \approx v_R$ due to the viewing angle.
These flows can clearly be seen in all of the SDW simulations and are clearest in the L-SDW simulation, where at $d_\odot \cos \ell \approx -4$ kpc, there is a 1.5 kpc by 8 kpc region of consistently negative $v_d$.
The presence of such a region is a basic and fundamental prediction of the SDW model --- without this flow, there would be no gaseous arm.

There is no such region of consistent inflow or outflow in the outer Galaxy portion of either KT map.
The lack of this region of consistent flow cannot be the result of assuming an incorrect Solar motion relative to the local standard of rest.
Adjusting the Solar motion changes the velocity field in a way that varies only with position on the sky, while the changes that would need to be made to the KT maps in order to introduce a consistent radial flow require a correction that varies with distance.

The velocity field around the Perseus arm HMSFRs in the 2nd quadrant is hard to reconcile with the SDW prediction that dense gas should be downstream from a rapid radial flow.
If the Perseus arm is inside corotation, there should be gas flowing outward towards the HMSFRs from the near side of the arm.
If it is instead outside corotation, there should be gas flowing inward towards the HMSFRs from the far side of the arm.
Instead, these HMSFRs are entrained in a strong inward flow.
There is no region that could be``feeding" the Perseus arm in the 2nd quadrant.

These two observations -- a lack of consistent radial flow across the 2nd and 3rd quadrants and the velocity field in the vicinity of the HMSFRs -- are based on large portions of both KT maps and directly contradict qualitative predictions of the SDW model.
The simulations we compare the KT map with are somewhat idealized.
In particular, most of the simulations do not include feedback and do not produce certain expected instabilities \citep[e.g.][]{2006MNRAS.367..873D}.
However, the effect of any missing physical processes would need to be strong enough to destroy large-scale properties of gaseous SDW spiral arms.
The SDW model with this missing physics included would need to be qualitatively different from the relatively simple SDW model.

The dynamic spiral structure model does not have as many broad predictions as the SDW model.
For example, there is no expectation of coherent flow over long spatial scales.
This is one reason for the greater diversity of $v_d$ fields in the dynamic spiral structure simulations relative to those in the SDW simulations.
The two main qualitative predictions of the dynamic spiral structure model are that dense gas in growing spiral arms should be located at the center of converging flows and that spiral arms should, at some point, dissipate \citep{2016MNRAS.460.2472B}.
When a spiral arm is in the dissipation phase, the gas in that arm should be located in diverging parts of the velocity field.

\begin{figure}
  \includegraphics{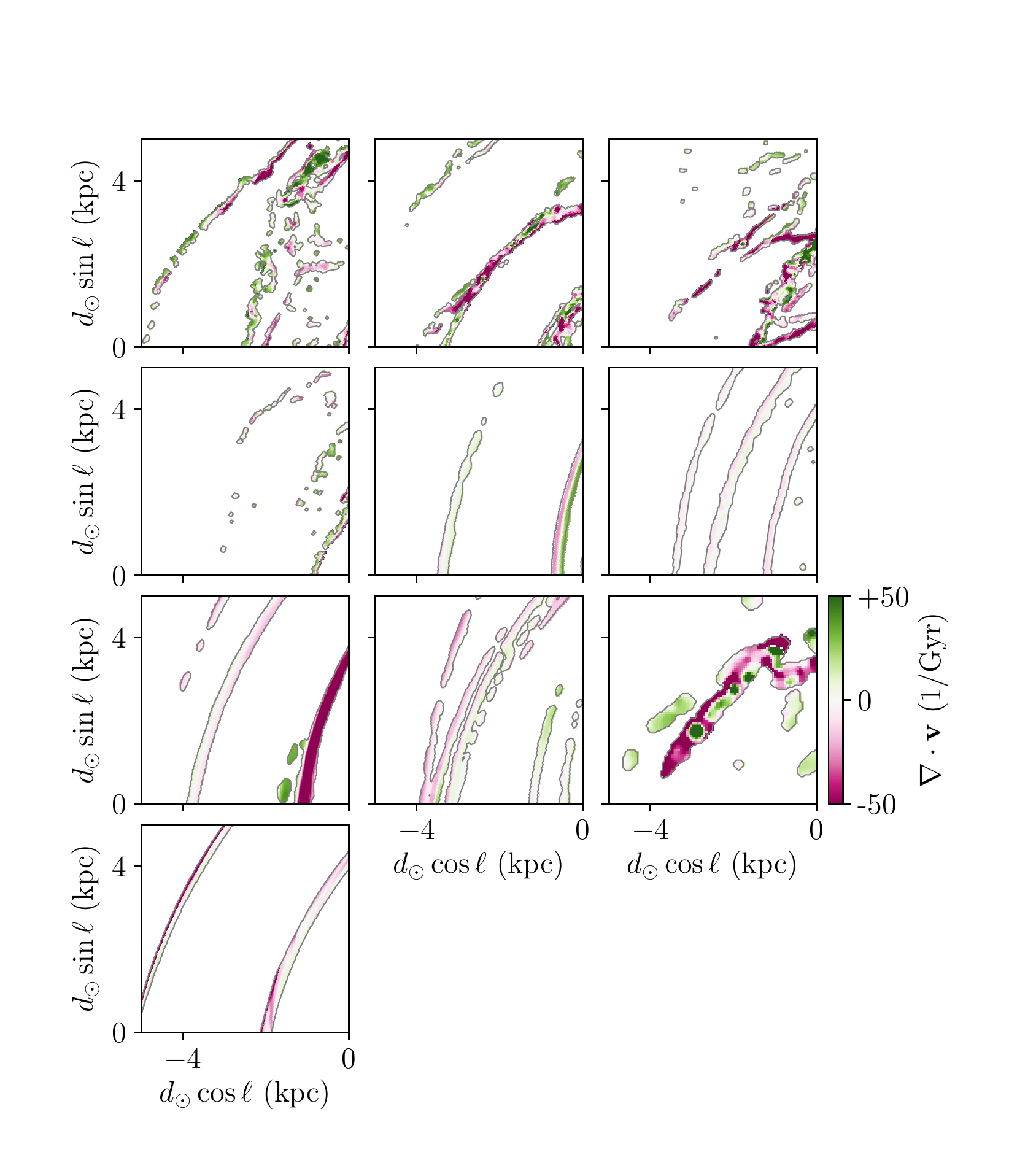}
  \caption{The divergence, ${\nabla}\cdot {\bf v}$, of the ISM velocity field of in dense regions in the spiral structure simulations. The divergence is indicated by the color scale. The regions shown are in the top surface density decile in their simulations. The simulations are, going first from left to right and then from top to bottom: P-D1 through P-D4, P-SDW1 through P-SDW 4, K-D, and L-SDW. Note that this figure, unlike the similarly structured Figures \ref{fig:models-and-KT}, \ref{fig:parameter-sequence-KT-maps}, and \ref{fig:model-density-vs-HMSFRs}, only shows the second quadrant of the Galactic plane. }
  \label{fig:divergences}
\end{figure}

Based on the divergence of the velocity field at the location of dense gas, which we show in Figure \ref{fig:divergences}, the Perseus-like arm in the P-D1 simulation is in the dissipation phase.
In the P-D2 and P-D3 simulations, the Perseus-like arms are located at sites of convergence while in the P-D4 simulation, there is no analog to the Perseus arm.
The fact that the Perseus-like arms are at a site of convergence in P-D2 and P-D3 can be seen directly from the $v_d$ maps, in which the Perseus-like arms are located at places where gas is converging along the line of sight.
Since the KT maps are mostly consistent with the P-D1 simulation but are not consistent with the P-D2, P-D3, or P-D4 simulations, we can conclude that the Milky Way has dynamic spiral structure and that the nearby section of the Perseus arm is in the process of dissipating.

It is true that the P-D1 simulation is not a perfect match for KT.
For example, in the P-D1 simulation, the region beyond the dense gas of the Perseus arm is flowing outward while in the KT map, this region has $v_d \approx 0$ km/s.
This and other small differences are not required by the dynamic spiral structure model and are instead realization-specific.
The outward flow outside the Perseus arm in the P-D1 simulation is gas converging on the Outer arm, which in the simulation is closer to the Sun than appears to be the case in the actual Milky Way.
The spacing between arms is not a fundamental prediction of the dynamic spiral structure model, as can be seen from the range of separations between arms in P-D1 through P-D4.
The velocity field we find using KT requires degrees of freedom in the velocity field that are not available in the SDW model but are available in the dynamic model.
Small, non-qualitative differences such as those between the P-D1 simulation and the KT maps are preferable to the broad and qualitative differences between the SDW model and the KT maps.

Our findings that the gaseous spiral arms in the outer galaxy are likely dynamic, and that the Perseus are is in a dissipating phase, are consistent with several recent analyses of stellar velocities based on \emph{Gaia} DR1 and DR2.
\citet{2018arXiv180510236Q} analyzed overdensities and boundaries between overdensities in the stellar velocity distribution seen in Gaia DR2 to determine the pattern speeds of nearby spiral arms.
Based on this analysis, the spiral arms near the Sun are all corotating or nearly corotating with the disk, which is inconsistent with the SDW model but predicted by the dynamic model.
\citet{2018arXiv180602832H} show that ridges in the same stellar velocity distribution resemble predictions from the dynamic model.
An analysis of Cepheid velocities based on Gaia DR1 data by \citet{Baba:2018ir} found that the Perseus arm is likely to be a dynamic arm  in the dissipation phase.

\begin{figure}
  \includegraphics{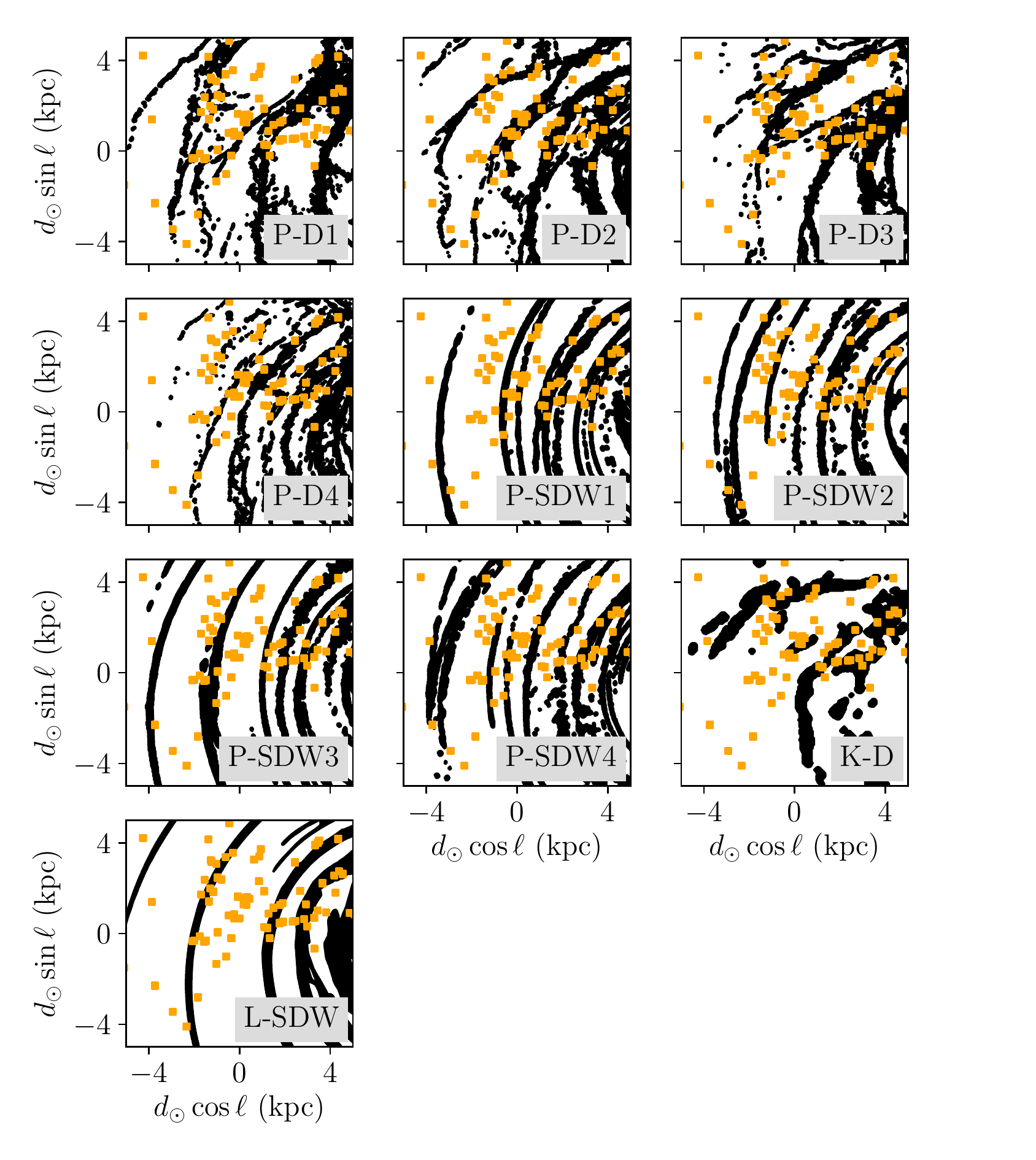}
  \caption{The location of ISM in the top surface density decile in each simulation is shown in black. The locations of the \citet{2014ApJ...783..130R} high mass star formation regions are shown in orange.}
  \label{fig:model-density-vs-HMSFRs}
\end{figure}

One potential counterargument to a dynamical model for the Milky Way's gaseous spiral structure is the ``tidiness'' of the HMSFR locations.
As R+14 have shown, one can assign most of the known HMSFRs to spiral arms whose pitch angles and lengths seem more consistent with the SDW model than the dynamical model.
In Figure \ref{fig:model-density-vs-HMSFRs}, we compare the location of dense gas in each of the simulations with the locations of the HMSFRs.
Based on this naive, by-eye comparison, we would argue that the distribution of HMSFRs is not so different from many of the arrangements of dense gas in dynamic simulations.
This interpretation can, for example, provide a simple explanation for the ``gap'' in the Perseus arm at roughly $\ell=180^\circ$ --- the HMSFRs in the second quadrant and the HMSFRs in the thrid quadrant actually belong to two different spiral arms.
There is not a perfect mapping between dense gas and HMSFRs and distance uncertainties tend to elongate structures along the line of sight.
Despite these caveats, we believe this comparison is sufficient to establish that the distribution of known HMSFRs does not rule out the possibility that the outer Milky Way has flocculent spiral arms with high pitch angles, and thus is consistent with dynamic spiral structure.

\section{Conclusion}\label{sec:conclusion}

In this work, we have been able to distinguish between different theories of spiral structure for the Milky Way using maps of the Milky Way ISM velocity field. In particular

\begin{itemize}
\item{In a process called \emph{Kinetic Tomography}, we have constructed a map of Milky Way's ISM velocity field using observations of a diffuse interstellar band ISM absorption line toward distant disk stars that is consistent with previous maps.}
\item{Spiral density wave theory and dynamic spiral theory make significantly different predictions about this velocity field, especially in the neighborhood of spiral arms.}
\item{We find that one simulation of each spiral theory has rough quantitative agreement with the maps of the Milky Way velocity field.}
\item{We find that only dynamic spiral theory can account for the Milky Way's complex velocity field, and the divergence of the velocity field detected at the Perseus Arm.}
\end{itemize}

This work has shown that measuring the velocity field of the dynamically cold interstellar medium can provide a unique and powerful tool for distinguishing between theories of how our Milky Way is structured. There are a number of upcoming programs that will continue to enhance our view of the velocity field of the ISM. APOGEE-II \citep{Zasowski_2017_apogee2targeting} will explore the southern hemisphere in DIBs, and the DECaPS program \citep{2018ApJS..234...39S}, designed to fill in the fourth quadrant of the \citet{2015ApJ...810...25G} 3D dust map, will allow us to construct all-sky G\&D maps. These maps will allow us to study the complete Scutum-Centaurus arm in the inner galaxy, home of much of Milky Way's star formation. SDSS-V, expected to begin observations in 2020, will observe many millions of giants across the Milky Way disk at APOGEE resolution, and has a subprogram devoted to measuring DIBs and dust toward stars within 4 kpc of the Sun. This will give us both better reach to study velocity fields across the entire Milky Way, and a much more detailed picture locally.

\acknowledgments
K.T. thanks Daisuke Kawata, Zhi Li, and Alex Petitt for providing simulation data and guidance on how to interpret it. K.T. and J.P. were supported by the National Science Foundation under grant No. 1616177. G.Z.\,acknowledges support from the Barry~M.~Lasker Data Science Research Fellowship, sponsored by the Space Telescope Science Institute in Baltimore, MD, USA.

Funding for SDSS-III has been provided by the Alfred P. Sloan Foundation, the Participating Institutions, the National Science Foundation, and the U.S. Department of Energy Office of Science. The SDSS-III web site is \url{http://www.sdss3.org/}.

SDSS-III is managed by the Astrophysical Research Consortium for the Participating Institutions of the SDSS-III Collaboration including the University of Arizona, the Brazilian Participation Group, Brookhaven National Laboratory, Carnegie Mellon University, University of Florida, the French Participation Group, the German Participation Group, Harvard University, the Instituto de Astrof\'isica de Canarias, the Michigan State/Notre Dame/JINA Participation Group, Johns Hopkins University, Lawrence Berkeley National Laboratory, Max Planck Institute for Astrophysics, Max Planck Institute for Extraterrestrial Physics, New Mexico State University, New York University, The Ohio State University, Pennsylvania State University, University of Portsmouth, Princeton University, the Spanish Participation Group, University of Tokyo, University of Utah, Vanderbilt University, University of Virginia, University of Washington, and Yale University.

Funding for the Sloan Digital Sky Survey IV has been provided by the Alfred P. Sloan Foundation, the U.S. Department of Energy Office of Science, and the Participating Institutions. SDSS-IV acknowledges support and resources from the Center for High-Performance Computing at the University of Utah. The SDSS web site is \url{www.sdss.org}.

SDSS-IV is managed by the Astrophysical Research Consortium for the Participating Institutions of the SDSS Collaboration including the Brazilian Participation Group, the Carnegie Institution for Science, Carnegie Mellon University, the Chilean Participation Group, the French Participation Group, Harvard-Smithsonian Center for Astrophysics, Instituto de Astrof\'isica de Canarias, The Johns Hopkins University, Kavli Institute for the Physics and Mathematics of the Universe (IPMU) / University of Tokyo, Lawrence Berkeley National Laboratory, Leibniz Institut f\"ur Astrophysik Potsdam (AIP),  Max-Planck-Institut f\"ur Astronomie (MPIA Heidelberg), Max-Planck-Institut f\"ur Astrophysik (MPA Garching), Max-Planck-Institut f\"ur Extraterrestrische Physik (MPE), National Astronomical Observatories of China, New Mexico State University, New York University, University of Notre Dame, Observat\'ario Nacional / MCTI, The Ohio State University, Pennsylvania State University, Shanghai Astronomical Observatory, United Kingdom Participation Group, Universidad Nacional Aut\'onoma de M\'exico, University of Arizona, University of Colorado Boulder, University of Oxford, University of Portsmouth, University of Utah, University of Virginia, University of Washington, University of Wisconsin, Vanderbilt University, and Yale University.

This work has made use of data from the European Space Agency (ESA) mission
{\it Gaia} (\url{https://www.cosmos.esa.int/gaia}), processed by the {\it Gaia}
Data Processing and Analysis Consortium (DPAC,
\url{https://www.cosmos.esa.int/web/gaia/dpac/consortium}). Funding for the DPAC
has been provided by national institutions, in particular the institutions
participating in the {\it Gaia} Multilateral Agreement.

\vspace{5mm}
\facilities{Sloan}

\software{astropy \citep{astropy_2013,astropy_2018},
emcee \citep{2013PASP..125..306F},
numpy \citep{vanderWalt:dp},
matplotlib \citep{2007CSE.....9...90H}
}

\appendix
\section{Marginalizing over profile amplitude and continuum offset}
\label{appendix:analytic-marginalization}
The integrals over profile amplitude $a$ and continuum offset $b$ in Equation \ref{eqn:pixel-likelihood} can be evaluated analytically when the values of the other parameters are held fixed.
Given a profile shape ${\bf f}(v_d, \sigma_v^2)$, an uncertainty $\sigma_y^2$, and prior variances $\sigma_a^2$ and $\sigma_b^2$ for $a$ and $b$, inferring $a$ and $b$ is a regularized linear problem.
Following e.g. \citet{2017RNAAS...1a...7L}, the integral with limits $-\infty$ and $+\infty$ for both $a$ and $b$ evaluates to
\begin{align}
\label{eqn:inf-marg-compact}
p\left({\bf y} \vert {\bf f}, \sigma_y^2, \sigma_a^2, \sigma_b^2\right) &=
\mathcal{N}({\bf y}; {\bf 0}, \sigma_y^2 I + A \Lambda A^T)\\
&= \frac{1}{\left(2\pi \right)^{N/2} \det\left(\sigma_y^2 I + A \Lambda A^T \right)}
\exp \left[ - \frac{1}{2}\,
{\bf y}^T \left(\sigma_y^2 I + A \Lambda A^T \right)^{-1} {\bf y}
\right]
\end{align}
Here, $I$ is the identity matrix, $A$ is the \emph{design matrix} for the problem, $\Lambda$ is a 2-by-2 diagonal matrix with diagonal entries $\sigma_a^2$ and $\sigma_b^2$, and $N$ is the length of ${\bf y}$.
The first column of $A$ consists of the elements of ${\bf f}$ and the second column consists of ones.

An explicit form for the determinant term can be obtained by invoking the generalized matrix determinant lemma.
We give the expression in terms of the inverse variances $\tau_y$, $\tau_a$, and $\tau_b$ for notational convenience:
\begin{align}
\det\left(\sigma_y^2 I + A \Lambda A^T \right) &=
\det\left(\Lambda^{-1} + \tau_y A^T I A \right) \det\left( \tau_y I \right) \det\left( \Lambda \right)\\
&= \frac{1}{\tau_a \tau_b \tau_y^N}
\left(
\left( \tau_a + \tau_y \sum_i f_i^2 \right) \left(\tau_b + N \tau_y \right) -
\left(\tau_y \sum_i f_i \right)^2
\right),
\end{align}
where all sums are from $i=1$ to $N$.
The explicit form of the argument of the exponential is unwieldy and incovenient to use.
In terms of the regularized least-squares solution $\hat{\bf x}$ to this linear problem and the corresponding least-squares prediction $A \hat{\bf x} \equiv \hat{\bf y}$, the argument is:
\begin{align}
{\bf y}^T \left(\sigma_y^2 I + A \Lambda A^T \right)^{-1} {\bf y} &= \tau_y {\bf y}^T I {\bf y} - \tau_y {\bf y}^T I \hat{\bf y}\\
&= \tau_y \left( \sum_i y_i^2 - \sum_i y_i \hat{y}_i \right).
\end{align}
The intermediate steps use the Woodbury identity and the definition of the regularized least-squares estimator.

When the integral over $b$ is still taken from $-\infty$ to $+\infty$ but the integral over $a$ is taken from $0$ to $+\infty$, as is done in the model we use to derive a prior on the DIB profile's width (Section \ref{subsec:sigmav-prior}), the result is the product of the expression given in Equation \ref{eqn:inf-marg-compact} and a number between 0 and 1.
To arrive at this result, we marginalize over $a$ and $b$ separately instead of simultaneously.
This can be done in our case because the prior on $a$ and $b$ assumes they are independent.
Marginalizing over $b$ gives
\begin{align}
p\left({\bf y} \vert {\bf f}, a, \sigma_y^2, \sigma_a^2, \sigma_b^2\right) &= \mathcal{N}({\bf y}; {\bf f} a, \sigma_y^2 I + \sigma_b {\bf 1 1}^T)\\
&\equiv \mathcal{N}({\bf y}; {\bf f} a, C_b)
\end{align}
where ${\bf 1}$ is an $N$-by-1 vector of ones.
Integrating this expression over $a$ from $-\infty$ to $+\infty$ gives
\begin{align}
p\left({\bf y} \vert {\bf f}, \sigma_y^2, \sigma_a^2, \sigma_b^2\right) &= \mathcal{N}({\bf y}; {\bf f} a, C_b + \sigma_b {\bf f f}^T)\\
&\equiv \mathcal{N}({\bf y}; {\bf f} a, \sigma_y^2 I + \sigma_a {\bf f f}^T + \sigma_b {\bf 1 1}^T),
\end{align}
which is equal to Equation \ref{eqn:inf-marg-compact}.

If we instead integrate over $a$ from 0 to $+\infty$, the only term that changes in Equation 12 of \citet{2017RNAAS...1a...7L} is the integral itself.
Using their variables $h$ and $\Sigma$,
\begin{align}
\int_0^{+\infty} \exp\left[ (a - h) \Sigma^{-1} (a-h) \right]\, {\rm d}a = 2\pi \Sigma \times
\frac{1}{2} \left(1 - {\rm erf}\left[\frac{-h}{\sqrt{2 \Sigma}} \right] \right).
\end{align}
The quantity $\Sigma$, which \citet{2017RNAAS...1a...7L} define in terms of its inverse in their Equation 9, evaluates to
\begin{align}
  \Sigma^{-1} &= \tau_a + \tau_y\sum_i f_i^2 - \frac{\left(\tau_y \sum_i f_i \right)^2}{\tau_b + N \tau_y}.
\end{align}
The quantity $h$, which is defined in their Equation 10, evaluates to
\begin{align}
  h &= \Sigma \left(\tau_y \sum_i y_i f_i - \frac{\tau_y^2 \left(\sum_i y_i \right) \left( \sum_i f_i \right)}{\tau_b + N \tau_y}  \right).
\end{align}
Multiplying Equation \ref{eqn:inf-marg-compact} by $\frac{1}{2} \left(1 - {\rm erf}\left[\frac{-h}{\sqrt{2 \Sigma}} \right] \right)$ gives the marginal likelihood of ${\bf y}$ when the DIB profile amplitude is constrained to be positive.

\bibliography{main.bbl}

\begin{thebibliography}{}
\expandafter\ifx\csname natexlab\endcsname\relax\def\natexlab#1{#1}\fi
\providecommand{\url}[1]{\href{#1}{#1}}

\bibitem[{{Abolfathi} {et~al.}(2018){Abolfathi}, {Aguado}, {Aguilar}, {Allende
  Prieto}, {Almeida}, {Ananna}, {Anders}, {Anderson}, {Andrews}, {Anguiano}, \&
  et~al.}]{Abolfathi_2017_SDSSDR14}
{Abolfathi}, B., {Aguado}, D.~S., {Aguilar}, G., {et~al.} 2018, \apjs, 235, 42

\bibitem[{{Astropy Collaboration} {et~al.}(2013){Astropy Collaboration},
  {Robitaille}, {Tollerud}, {Greenfield}, {Droettboom}, {Bray}, {Aldcroft},
  {Davis}, {Ginsburg}, {Price-Whelan}, {Kerzendorf}, {Conley}, {Crighton},
  {Barbary}, {Muna}, {Ferguson}, {Grollier}, {Parikh}, {Nair}, {Unther},
  {Deil}, {Woillez}, {Conseil}, {Kramer}, {Turner}, {Singer}, {Fox}, {Weaver},
  {Zabalza}, {Edwards}, {Azalee Bostroem}, {Burke}, {Casey}, {Crawford},
  {Dencheva}, {Ely}, {Jenness}, {Labrie}, {Lian Lim}, {Pierfederici},
  {Pontzen}, {Ptak}, {Refsdal}, {Servillat}, \& {Streicher}}]{astropy_2013}
{Astropy Collaboration}, {Robitaille}, T.~P., {Tollerud}, E.~J., {et~al.} 2013,
  \aap, 558, A33

\bibitem[{Baba {et~al.}(2018)Baba, Kawata, Matsunaga, Grand, \&
  Hunt}]{Baba:2018ir}
Baba, J., Kawata, D., Matsunaga, N., Grand, R. J.~J., \& Hunt, J. A.~S. 2018,
  ApJ, 853, L23

\bibitem[{Baba {et~al.}(2016)Baba, Morokuma-Matsui, Miyamoto, Egusa, \&
  Kuno}]{2016MNRAS.460.2472B}
Baba, J., Morokuma-Matsui, K., Miyamoto, Y., Egusa, F., \& Kuno, N. 2016,
  \mnras, 460, 2472

\bibitem[{{Blanton} {et~al.}(2017){Blanton}, {Bershady}, {Abolfathi},
  {Albareti}, {Allende Prieto}, {Almeida}, {Alonso-Garc{\'{\i}}a}, {Anders},
  {Anderson}, {Andrews}, \& et~al.}]{Blanton_2017_sdss4}
{Blanton}, M.~R., {Bershady}, M.~A., {Abolfathi}, B., {et~al.} 2017, \aj, 154,
  28

\bibitem[{{Bovy} {et~al.}(2014){Bovy}, {Nidever}, {Rix}, {Girardi}, {Zasowski},
  {Chojnowski}, {Holtzman}, {Epstein}, {Frinchaboy}, {Hayden}, {Rodrigues},
  {Majewski}, {Johnson}, {Pinsonneault}, {Stello}, {Allende Prieto}, {Andrews},
  {Basu}, {Beers}, {Bizyaev}, {Burton}, {Chaplin}, {Cunha}, {Elsworth},
  {Garc{\'{\i}}a}, {Garc{\'{\i}}a-Her{\'n}andez}, {Garc{\'{\i}}a P{\'e}rez},
  {Hearty}, {Hekker}, {Kallinger}, {Kinemuchi}, {Koesterke},
  {M{\'e}sz{\'a}ros}, {Mosser}, {O'Connell}, {Oravetz}, {Pan}, {Robin},
  {Schiavon}, {Schneider}, {Schultheis}, {Serenelli}, {Shetrone}, {Silva
  Aguirre}, {Simmons}, {Skrutskie}, {Smith}, {Stassun}, {Weinberg}, {Wilson},
  \& {Zamora}}]{Bovy_2014_APOGEE_RC_catalog}
{Bovy}, J., {Nidever}, D.~L., {Rix}, H.-W., {et~al.} 2014, \apj, 790, 127

\bibitem[{{Casey} {et~al.}(2016){Casey}, {Hogg}, {Ness}, {Rix}, {Ho}, \&
  {Gilmore}}]{Casey_2016_cannon2}
{Casey}, A.~R., {Hogg}, D.~W., {Ness}, M., {et~al.} 2016, arXiv:1603.03040

\bibitem[{Cohen {et~al.}(1980)Cohen, Cong, Dame, \&
  Thaddeus}]{1980ApJ...239L..53C}
Cohen, R.~S., Cong, H., Dame, T.~M., \& Thaddeus, P. 1980, ApJ, 239, L53

\bibitem[{{Cox} {et~al.}(2014){Cox}, {Cami}, {Kaper}, {Ehrenfreund}, {Foing},
  {Ochsendorf}, {van Hooff}, \& {Salama}}]{Cox_2014_xshooterDIBs}
{Cox}, N.~L.~J., {Cami}, J., {Kaper}, L., {et~al.} 2014, \aap, 569, A117

\bibitem[{{Dame} \& {Thaddeus}(2011)}]{Dame_2011_outerspiralarm}
{Dame}, T.~M., \& {Thaddeus}, P. 2011, \apjl, 734, L24

\bibitem[{Dobbs \& Baba(2014)}]{2014PASA...31...35D}
Dobbs, C., \& Baba, J. 2014, \pasa, 31, e035

\bibitem[{Dobbs \& Bonnell(2006)}]{2006MNRAS.367..873D}
Dobbs, C.~L., \& Bonnell, I.~A. 2006, \mnras, 367, 873

\bibitem[{D'Onghia {et~al.}(2013)D'Onghia, Vogelsberger, \&
  Hernquist}]{2013ApJ...766...34D}
D'Onghia, E., Vogelsberger, M., \& Hernquist, L. 2013, ApJ, 766, 34

\bibitem[{{Eisenstein} {et~al.}(2011){Eisenstein}, {Weinberg}, {Agol},
  {Aihara}, {Allende Prieto}, {Anderson}, {Arns}, {Aubourg}, {Bailey},
  {Balbinot}, \& et~al.}]{Eisenstein_11_sdss3overview}
{Eisenstein}, D.~J., {Weinberg}, D.~H., {Agol}, E., {et~al.} 2011, \aj, 142, 72

\bibitem[{{Elyajouri} {et~al.}(2017){Elyajouri}, {Cox}, \&
  {Lallement}}]{Elyajouri_2017_DIBinB68}
{Elyajouri}, M., {Cox}, N.~L.~J., \& {Lallement}, R. 2017, \aap, 605, L10

\bibitem[{{Elyajouri} {et~al.}(2016){Elyajouri}, {Monreal-Ibero}, {Remy}, \&
  {Lallement}}]{Elyajouri_2016_apogeetelluricDIBs}
{Elyajouri}, M., {Monreal-Ibero}, A., {Remy}, Q., \& {Lallement}, R. 2016,
  \apjs, 225, 19

\bibitem[{Foreman-Mackey {et~al.}(2013)Foreman-Mackey, Hogg, Lang, \&
  Goodman}]{2013PASP..125..306F}
Foreman-Mackey, D., Hogg, D.~W., Lang, D., \& Goodman, J. 2013, \pasp, 125, 306

\bibitem[{{Gaia Collaboration} {et~al.}(2018){Gaia Collaboration}, {Brown},
  {Vallenari}, {Prusti}, {de Bruijne}, {Babusiaux}, \&
  {Bailer-Jones}}]{GaiaCollab_2018_gaiaDR2}
{Gaia Collaboration}, {Brown}, A.~G.~A., {Vallenari}, A., {et~al.} 2018,
  arXiv:1804.09365

\bibitem[{{Gaia Collaboration} {et~al.}(2016){Gaia Collaboration}, {Prusti},
  {de Bruijne}, {Brown}, {Vallenari}, {Babusiaux}, {Bailer-Jones}, {Bastian},
  {Biermann}, {Evans}, \& et~al.}]{GaiaCollab_2016_gaia}
{Gaia Collaboration}, {Prusti}, T., {de Bruijne}, J.~H.~J., {et~al.} 2016,
  \aap, 595, A1

\bibitem[{{Garc{\'{\i}}a P{\'e}rez} {et~al.}(2016){Garc{\'{\i}}a P{\'e}rez},
  {Allende Prieto}, {Holtzman}, {Shetrone}, {M{\'e}sz{\'a}ros}, {Bizyaev},
  {Carrera}, {Cunha}, {Garc{\'{\i}}a-Hern{\'a}ndez}, {Johnson}, {Majewski},
  {Nidever}, {Schiavon}, {Shane}, {Smith}, {Sobeck}, {Troup}, {Zamora},
  {Weinberg}, {Bovy}, {Eisenstein}, {Feuillet}, {Frinchaboy}, {Hayden},
  {Hearty}, {Nguyen}, {O'Connell}, {Pinsonneault}, {Wilson}, \&
  {Zasowski}}]{GarciaPerez_2016_aspcap}
{Garc{\'{\i}}a P{\'e}rez}, A.~E., {Allende Prieto}, C., {Holtzman}, J.~A.,
  {et~al.} 2016, \aj, 151, 144

\bibitem[{{Geballe} {et~al.}(2011){Geballe}, {Najarro}, {Figer},
  {Schlegelmilch}, \& {de La Fuente}}]{Geballe_2011_IRdibs}
{Geballe}, T.~R., {Najarro}, F., {Figer}, D.~F., {Schlegelmilch}, B.~W., \& {de
  La Fuente}, D. 2011, \nat, 479, 200

\bibitem[{Green {et~al.}(2015)Green, Schlafly, Finkbeiner, Rix, Martin,
  Burgett, Draper, Flewelling, Hodapp, Kaiser, Kudritzki, Magnier, Metcalfe,
  Price, Tonry, \& Wainscoat}]{2015ApJ...810...25G}
Green, G.~M., Schlafly, E.~F., Finkbeiner, D.~P., {et~al.} 2015, ApJ, 810, 25

\bibitem[{{Holtzman} {et~al.}(2015){Holtzman}, {Shetrone}, {Johnson}, {Allende
  Prieto}, {Anders}, {Andrews}, {Beers}, {Bizyaev}, {Blanton}, {Bovy},
  {Carrera}, {Chojnowski}, {Cunha}, {Eisenstein}, {Feuillet}, {Frinchaboy},
  {Galbraith-Frew}, {Garc{\'{\i}}a P{\'e}rez}, {Garc{\'{\i}}a-Hern{\'a}ndez},
  {Hasselquist}, {Hayden}, {Hearty}, {Ivans}, {Majewski}, {Martell},
  {Meszaros}, {Muna}, {Nidever}, {Nguyen}, {O'Connell}, {Pan}, {Pinsonneault},
  {Robin}, {Schiavon}, {Shane}, {Sobeck}, {Smith}, {Troup}, {Weinberg},
  {Wilson}, {Wood-Vasey}, {Zamora}, \& {Zasowski}}]{Holtzman_2015_apogeedata}
{Holtzman}, J.~A., {Shetrone}, M., {Johnson}, J.~A., {et~al.} 2015, \aj, 150,
  148

\bibitem[{Hunt {et~al.}(2018)Hunt, Hong, Bovy, Kawata, \&
  Grand}]{2018arXiv180602832H}
Hunt, J. A.~S., Hong, J., Bovy, J., Kawata, D., \& Grand, R. J.~J. 2018,
  arXiv:1806.02832

\bibitem[{Hunter(2007)}]{2007CSE.....9...90H}
Hunter, J.~D. 2007, CSE, 9, 90

\bibitem[{Kawata {et~al.}(2014)Kawata, Hunt, Grand, Pasetto, \&
  Cropper}]{Kawata:2014ho}
Kawata, D., Hunt, J. A.~S., Grand, R. J.~J., Pasetto, S., \& Cropper, M. 2014,
  \mnras, 443, 2757

\bibitem[{Li {et~al.}(2016)Li, Gerhard, Shen, Portail, \& Wegg}]{Li:2016dx}
Li, Z., Gerhard, O., Shen, J., Portail, M., \& Wegg, C. 2016, ApJ, 824, 1

\bibitem[{Lin \& Shu(1964)}]{1964ApJ...140..646L}
Lin, C.~C., \& Shu, F.~H. 1964, ApJ, 140, 646

\bibitem[{Luger {et~al.}(2017)Luger, Foreman-Mackey, \&
  Hogg}]{2017RNAAS...1a...7L}
Luger, R., Foreman-Mackey, D., \& Hogg, D.~W. 2017, RNAAS, 1, 7

\bibitem[{Luri {et~al.}(2018)Luri, A~Brown, Sarro, Arenou, Bailer-Jones,
  Castro-Ginard, de~Bruijne, Prusti, Babusiaux, \& Delgado}]{Luri:2018eu}
Luri, X., A~Brown, A.~G., Sarro, L., {et~al.} 2018, A{\&}A, in press

\bibitem[{{Majewski} {et~al.}(2017){Majewski}, {Schiavon}, {Frinchaboy},
  {Allende Prieto}, {Barkhouser}, {Bizyaev}, {Blank}, {Brunner}, {Burton},
  {Carrera}, {Chojnowski}, {Cunha}, {Epstein}, {Fitzgerald}, {Garc{\'{\i}}a
  P{\'e}rez}, {Hearty}, {Henderson}, {Holtzman}, {Johnson}, {Lam}, {Lawler},
  {Maseman}, {M{\'e}sz{\'a}ros}, {Nelson}, {Nguyen}, {Nidever}, {Pinsonneault},
  {Shetrone}, {Smee}, {Smith}, {Stolberg}, {Skrutskie}, {Walker}, {Wilson},
  {Zasowski}, {Anders}, {Basu}, {Beland}, {Blanton}, {Bovy}, {Brownstein},
  {Carlberg}, {Chaplin}, {Chiappini}, {Eisenstein}, {Elsworth}, {Feuillet},
  {Fleming}, {Galbraith-Frew}, {Garc{\'{\i}}a}, {Garc{\'{\i}}a-Hern{\'a}ndez},
  {Gillespie}, {Girardi}, {Gunn}, {Hasselquist}, {Hayden}, {Hekker}, {Ivans},
  {Kinemuchi}, {Klaene}, {Mahadevan}, {Mathur}, {Mosser}, {Muna}, {Munn},
  {Nichol}, {O'Connell}, {Parejko}, {Robin}, {Rocha-Pinto}, {Schultheis},
  {Serenelli}, {Shane}, {Silva Aguirre}, {Sobeck}, {Thompson}, {Troup},
  {Weinberg}, \& {Zamora}}]{Majewski_2017_apogeeoverview}
{Majewski}, S.~R., {Schiavon}, R.~P., {Frinchaboy}, P.~M., {et~al.} 2017, \aj,
  154, 94

\bibitem[{Marrese {et~al.}(2018)Marrese, Marinoni, Fabrizio, \&
  Altavilla}]{Marrese_2018_gaiaXmatch}
Marrese, P., Marinoni, S., Fabrizio, M., \& Altavilla, G. 2018, A\&A

\bibitem[{{McMillan} {et~al.}(2018){McMillan}, {Kordopatis}, {Kunder},
  {Binney}, {Wojno}, {Zwitter}, {Steinmetz}, {Bland-Hawthorn}, {Gibson},
  {Gilmore}, {Grebel}, {Helmi}, {Munari}, {Navarro}, {Parker}, {Seabroke},
  {Watson}, \& {Wyse}}]{McMillan_2018_tgas+rave}
{McMillan}, P.~J., {Kordopatis}, G., {Kunder}, A., {et~al.} 2018, \mnras, 477,
  5279

\bibitem[{Morgan {et~al.}(1952)Morgan, Sharpless, \&
  Osterbrock}]{Morgan:1952gm}
Morgan, W.~W., Sharpless, S., \& Osterbrock, D. 1952, \aj, 57, 3

\bibitem[{{Ness} {et~al.}(2015){Ness}, {Hogg}, {Rix}, {Ho}, \&
  {Zasowski}}]{Ness_2015_Cannon}
{Ness}, M., {Hogg}, D.~W., {Rix}, H.-W., {Ho}, A.~Y.~Q., \& {Zasowski}, G.
  2015, \apj, 808, 16

\bibitem[{{Nidever} {et~al.}(2015){Nidever}, {Holtzman}, {Allende Prieto},
  {Beland}, {Bender}, {Bizyaev}, {Burton}, {Desphande}, {Fleming},
  {Garc{\'{\i}}a P{\'e}rez}, {Hearty}, {Majewski}, {M{\'e}sz{\'a}ros}, {Muna},
  {Nguyen}, {Schiavon}, {Shetrone}, {Skrutskie}, {Sobeck}, \&
  {Wilson}}]{Nidever_2015_apogeereduction}
{Nidever}, D.~L., {Holtzman}, J.~A., {Allende Prieto}, C., {et~al.} 2015, \aj,
  150, 173

\bibitem[{Pettitt {et~al.}(2015)Pettitt, Dobbs, Acreman, \&
  Bate}]{Pettitt:2015kx}
Pettitt, A.~R., Dobbs, C.~L., Acreman, D.~M., \& Bate, M.~R. 2015, \mnras, 449,
  3911

\bibitem[{Pettitt {et~al.}(2014)Pettitt, Dobbs, Acreman, \&
  Price}]{Pettitt:2014ep}
Pettitt, A.~R., Dobbs, C.~L., Acreman, D.~M., \& Price, D.~J. 2014, \mnras,
  444, 919

\bibitem[{Portail {et~al.}(2015)Portail, Wegg, Gerhard, \&
  Martinez-Valpuesta}]{2015MNRAS.448..713P}
Portail, M., Wegg, C., Gerhard, O., \& Martinez-Valpuesta, I. 2015, \mnras,
  448, 713

\bibitem[{Quillen {et~al.}(2018)Quillen, Carrillo, Anders, McMillan, Hilmi,
  Monari, Minchev, Chiappini, Khalatyan, \& Steinmetz}]{2018arXiv180510236Q}
Quillen, A.~C., Carrillo, I., Anders, F., {et~al.} 2018, arXiv:1805.10236

\bibitem[{Reid {et~al.}(2014)Reid, Menten, Brunthaler, Zheng, Dame, Xu, Wu,
  Zhang, Sanna, Sato, Hachisuka, Choi, Immer, Moscadelli, Rygl, \&
  Bartkiewicz}]{2014ApJ...783..130R}
Reid, M.~J., Menten, K.~M., Brunthaler, A., {et~al.} 2014, ApJ, 783, 130

\bibitem[{Roberts(1969)}]{1969ApJ...158..123R}
Roberts, W.~W. 1969, ApJ, 158, 123

\bibitem[{{Salgado} {et~al.}(2017){Salgado}, {Gonz{\'a}lez-N{\'u}{\~n}ez},
  {Guti{\'e}rrez-S{\'a}nchez}, {Segovia}, {Dur{\'a}n}, {Hern{\'a}ndez}, \&
  {Arviset}}]{Salgado_2017_gaiaarchive}
{Salgado}, J., {Gonz{\'a}lez-N{\'u}{\~n}ez}, J., {Guti{\'e}rrez-S{\'a}nchez},
  R., {et~al.} 2017, A\&C, 21, 22

\bibitem[{{Santiago} {et~al.}(2016){Santiago}, {Brauer}, {Anders}, {Chiappini},
  {Queiroz}, {Girardi}, {Rocha-Pinto}, {Balbinot}, {da Costa}, {Maia},
  {Schultheis}, {Steinmetz}, {Miglio}, {Montalb{\'a}n}, {Schneider}, {Beers},
  {Frinchaboy}, {Lee}, \& {Zasowski}}]{Santiago_2016_apogeedistances}
{Santiago}, B.~X., {Brauer}, D.~E., {Anders}, F., {et~al.} 2016, \aap, 585, A42

\bibitem[{{Schlafly} \& {Finkbeiner}(2011)}]{Schlafly_2011_calibSFD}
{Schlafly}, E.~F., \& {Finkbeiner}, D.~P. 2011, \apj, 737, 103

\bibitem[{{Schlafly} {et~al.}(2018){Schlafly}, {Green}, {Lang}, {Daylan},
  {Finkbeiner}, {Lee}, {Meisner}, {Schlegel}, \&
  {Valdes}}]{2018ApJS..234...39S}
{Schlafly}, E.~F., {Green}, G.~M., {Lang}, D., {et~al.} 2018, \apjs, 234, 39

\bibitem[{Sellwood \& Carlberg(1984)}]{1984ApJ...282...61S}
Sellwood, J.~A., \& Carlberg, R.~G. 1984, ApJ, 282, 61

\bibitem[{Shu(2016)}]{Shu:2016ij}
Shu, F.~H. 2016, \araa, 54, 667

\bibitem[{Sofue(2012)}]{2012PASJ...64...75S}
Sofue, Y. 2012, \pasj, 64, 75

\bibitem[{{Tchernyshyov} \& {Peek}(2017)}]{Tchernyshyov_2017_kt1}
{Tchernyshyov}, K., \& {Peek}, J.~E.~G. 2017, \aj, 153, 8

\bibitem[{{The Astropy Collaboration} {et~al.}(2018){The Astropy
  Collaboration}, {Price-Whelan}, {Sip{\H o}cz}, {G{\"u}nther}, {Lim},
  {Crawford}, {Conseil}, {Shupe}, {Craig}, {Dencheva}, {Ginsburg},
  {VanderPlas}, {Bradley}, {P{\'e}rez-Su{\'a}rez}, {de Val-Borro}, {Aldcroft},
  {Cruz}, {Robitaille}, {Tollerud}, {Ardelean}, {Babej}, {Bachetti}, {Bakanov},
  {Bamford}, {Barentsen}, {Barmby}, {Baumbach}, {Berry}, {Biscani}, {Boquien},
  {Bostroem}, {Bouma}, {Brammer}, {Bray}, {Breytenbach}, {Buddelmeijer},
  {Burke}, {Calderone}, {Cano Rodr{\'{\i}}guez}, {Cara}, {Cardoso},
  {Cheedella}, {Copin}, {Crichton}, {D{\'A}vella}, {Deil}, {Depagne},
  {Dietrich}, {Donath}, {Droettboom}, {Earl}, {Erben}, {Fabbro}, {Ferreira},
  {Finethy}, {Fox}, {Garrison}, {Gibbons}, {Goldstein}, {Gommers}, {Greco},
  {Greenfield}, {Groener}, {Grollier}, {Hagen}, {Hirst}, {Homeier}, {Horton},
  {Hosseinzadeh}, {Hu}, {Hunkeler}, {Ivezi{\'c}}, {Jain}, {Jenness}, {Kanarek},
  {Kendrew}, {Kern}, {Kerzendorf}, {Khvalko}, {King}, {Kirkby}, {Kulkarni},
  {Kumar}, {Lee}, {Lenz}, {Littlefair}, {Ma}, {Macleod}, {Mastropietro},
  {McCully}, {Montagnac}, {Morris}, {Mueller}, {Mumford}, {Muna}, {Murphy},
  {Nelson}, {Nguyen}, {Ninan}, {N{\"o}the}, {Ogaz}, {Oh}, {Parejko}, {Parley},
  {Pascual}, {Patil}, {Patil}, {Plunkett}, {Prochaska}, {Rastogi}, {Reddy
  Janga}, {Sabater}, {Sakurikar}, {Seifert}, {Sherbert}, {Sherwood-Taylor},
  {Shih}, {Sick}, {Silbiger}, {Singanamalla}, {Singer}, {Sladen}, {Sooley},
  {Sornarajah}, {Streicher}, {Teuben}, {Thomas}, {Tremblay}, {Turner},
  {Terr{\'o}n}, {van Kerkwijk}, {de la Vega}, {Watkins}, {Weaver}, {Whitmore},
  {Woillez}, \& {Zabalza}}]{astropy_2018}
{The Astropy Collaboration}, {Price-Whelan}, A.~M., {Sip{\H o}cz}, B.~M.,
  {et~al.} 2018, arXiv:1801.02634

\bibitem[{van~de Hulst {et~al.}(1954)van~de Hulst, Muller, \&
  Oort}]{1954BAN....12..117V}
van~de Hulst, H.~C., Muller, C.~A., \& Oort, J.~H. 1954, BAN, 12, 117

\bibitem[{van~der Walt {et~al.}(2011)van~der Walt, Colbert, \&
  Varoquaux}]{vanderWalt:dp}
van~der Walt, S., Colbert, S.~C., \& Varoquaux, G. 2011, CSE, 13, 22

\bibitem[{{Wang} {et~al.}(2016){Wang}, {Shi}, {Pan}, {Chen}, {Zhao}, \&
  {Wicker}}]{Wang_2016_APOGEEdistances}
{Wang}, J., {Shi}, J., {Pan}, K., {et~al.} 2016, \mnras, 460, 3179

\bibitem[{Xu {et~al.}(2018)Xu, Bian, Reid, Li, Zhang, Yan, Dame, Menten, He,
  Liao, \& Tang}]{Xu:2018kg}
Xu, Y., Bian, S.~B., Reid, M.~J., {et~al.} 2018, A{\&}A, in press

\bibitem[{{Zasowski} {et~al.}(2013){Zasowski}, {Johnson}, {Frinchaboy},
  {Majewski}, {Nidever}, {Rocha Pinto}, {Girardi}, {Andrews}, {Chojnowski},
  {Cudworth}, {Jackson}, {Munn}, {Skrutskie}, {Beaton}, {Blake}, {Covey},
  {Deshpande}, {Epstein}, {Fabbian}, {Fleming}, {Garcia Hernandez}, {Herrero},
  {Mahadevan}, {M{\'e}sz{\'a}ros}, {Schultheis}, {Sellgren}, {Terrien}, {van
  Saders}, {Allende Prieto}, {Bizyaev}, {Burton}, {Cunha}, {da Costa},
  {Hasselquist}, {Hearty}, {Holtzman}, {Garc{\'{\i}}a P{\'e}rez}, {Maia},
  {O'Connell}, {O'Donnell}, {Pinsonneault}, {Santiago}, {Schiavon}, {Shetrone},
  {Smith}, \& {Wilson}}]{Zasowski_2013_apogeetargeting}
{Zasowski}, G., {Johnson}, J.~A., {Frinchaboy}, P.~M., {et~al.} 2013, \aj, 146,
  81

\bibitem[{Zasowski {et~al.}(2015)Zasowski, M{\'e}nard, Bizyaev,
  Garc{\'\i}a-Hern{\'a}ndez, P{\'e}rez, Hayden, Holtzman, Johnson, Kinemuchi,
  Majewski, Nidever, Shetrone, \& Wilson}]{Zasowski:2015hi}
Zasowski, G., M{\'e}nard, B., Bizyaev, D., {et~al.} 2015, ApJ, 798, 35

\bibitem[{Zasowski {et~al.}(2017)Zasowski, Cohen, Chojnowski, Santana, Oelkers,
  Andrews, Beaton, Bender, Bird, Bovy, Carlberg, Covey, Cunha, Dell'Agli,
  Fleming, Frinchaboy, Garc\'{i}a-Hern\'{a}ndez, Harding, Holtzman, Johnson,
  Kollmeier, Majewski, M\'{e}sz\'{a}ros, Munn, Munoz, Ness, Nidever, Poleski,
  Rom\'{a}n-Z\'{u}niga, Shetrone, Simon, Smith, Sobeck, Stringfellow,
  Szigeti\'{a}ros, Tayar, \& Troup}]{Zasowski_2017_apogee2targeting}
Zasowski, G., Cohen, R.~E., Chojnowski, S.~D., {et~al.} 2017, AJ, 154, 198

\end{thebibliography}

\end{document}